\documentclass[]{aa}
\usepackage[colorlinks,
            linkcolor=blue,       
            anchorcolor=blue, 
            citecolor=blue
            ]{hyperref}
\usepackage{amsmath}
\usepackage{booktabs}
\usepackage[table]{xcolor}
\usepackage{placeins}
\usepackage{float}
\definecolor{myblue}{RGB}{159, 217, 255} % 浅蓝色
\usepackage{colortbl}
\usepackage{tabularx}
\usepackage{natbib}
\usepackage{multirow}
\usepackage{subcaption} % 确保引入这个宏包
\bibpunct{(}{)}{;}{a}{}{,} % to follow the A&A style
\usepackage{graphicx}
\usepackage{enumitem}
\usepackage{txfonts}
\usepackage[normalem]{ulem} % normalem 防止改变 \emph 的样式
\usepackage{orcidlink}

\begin{document}

\title{SolarZip: An efficient and adaptive compression framework for Solar EUV imaging data}  \subtitle{Application to Solar Orbiter/EUI data}   

\author{
          Zedong Liu\inst{1,2,5}\thanks{These authors contributed equally to this work.}
          \and
          Song Tan\inst{3,4}\textsuperscript{$\star$}
          \and
          Alexander Warmuth\inst{3}
          \and
          Frédéric Schuller\inst{3}
          \and
          Yun Hong\inst{6}
          \and
          Wenjing Huang\inst{1,5}
          \and\\
          Yida Gu\inst{1,5}
          \and
          Bojing Zhu\inst{7,8,5}
          \and
          Guangming Tan\inst{1,5}
          \and
          Dingwen Tao\inst{1,5}\thanks{Corresponding author; taodingwen@ict.ac.cn}
          }
   \institute{Institute of Computing Technology, Chinese Academy of Sciences, Beijing 100190, China
   \and University of Electronic Science and Technology of China, Chengdu 610054, China
   \and Leibniz-Institut für Astrophysik Potsdam (AIP), An der Sternwarte 16, 14482 Potsdam, Germany
   \and Institut für Physik und Astronomie, Universität Potsdam, Karl-Liebknecht-Straße 24/25, 14476 Potsdam, Germany
   \and University of Chinese Academy of Sciences, Beijing 100049, China
   \and Minzu University of China, Beijing 100081, China
   \and Yunnan Observatories, Chinese Academy of Sciences, Kunming 650216, China
   \and Centre for Astronomical Mega-Science, Chinese Academy of Sciences, Beijing 100012, China}

   \date{Received/ accepted}

% \abstract{}{}{}{}{}
% 5 {} token are mandatory
 
  \abstract
  % context heading (optional)
    {With the advancement of solar physics research, next-generation solar space missions and ground-based telescopes face significant challenges in efficiently transmitting and/or storing large-scale observational data.}
 % aims heading (mandatory)
    {We have developed an efficient compression and evaluation framework for solar EUV data, specifically optimized for Solar Orbiter Extreme Ultraviolet Imager (EUI) data. It significantly reduces the data volume, while preserving scientific usability.}
% methods heading (mandatory)
    {We evaluated four error-bounded lossy compressors across two Solar Orbiter/EUI datasets spanning 50 months of observations. However,  existing methods
cannot perfectly handle the EUV data. Building on this analysis, we developed SolarZip, an adaptive compression framework featuring: (1) a hybrid strategy controller that dynamically selects the optimal compression strategy; (2) enhanced spline interpolation predictors with grid-wise anchor points and level-wise error bound auto-tuning; and (3) a comprehensive two-stage evaluation methodology integrating standard distortion metrics with domain-specific post hoc scientific analyses.}
% results heading (mandatory)
    {Our  SolarZip framework achieved a data compression ratio of up to 800× for Full Sun Imager (FSI) data and 500× for High Resolution Imager (HRI$_{\text{EUV}}$) data. It significantly outperformed both traditional and advanced algorithms, achieving 3-50× higher compression ratios than traditional algorithms, surpassing the second-best algorithm by up to 30\%. Simulation experiments verified that SolarZip can reduce data transmission time by up to 270,× while ensuring the preservation of scientific usability.} 
% conclusions heading (optional)
    {The SolarZip framework significantly enhances solar observational data compression efficiency, while preserving scientific usability by dynamically selecting the optimal compression methods based on observational scenarios and user requirements. This approach offers a promising data management solution for deep space missions, such as Solar Orbiter.}


\maketitle

\section{Introduction}
With the advancement of solar physics research, next-generation solar space missions and ground-based telescopes demand increasingly higher spatial and temporal resolutions for observational data. This has made the efficient transmission and processing of large-scale data an urgent challenge. Solar Orbiter \citep{2020A&A...642A...1M}, a collaborative mission between the European Space Agency (ESA) and National Aeronautics and Space Administration (NASA), was successfully launched in February 2020. Its unique orbital design enables unprecedented close-up observations of the Sun (approaching as near as 0.28 AU) and provides the first high-resolution images of the solar polar regions. However, due to inherent telemetry constraints of deep space missions, the amount of data  observed by Solar Orbiter surpasses its transmission capabilities by a substantial margin, making efficient on-board data compression essential for achieving the mission's scientific objectives \citep{fischer2017jpeg2000}. The Extreme Ultraviolet Imager (EUI, \cite{rochus2020solar}), one of Solar Orbiter's core remote sensing instruments, consists of a Full Sun Imager (FSI) and two High Resolution Imagers (HRI$_{\text{EUV}}$ and HRI$_{\text{Ly}\alpha}$), providing comprehensive observations from the chromosphere to the corona in EUV wavelengths (17.4 nm, 30.4 nm) and the Lyman-$\alpha$ band. EUI data exhibit significant high dynamic range characteristics, with intensity differences between bright and dark regions spanning several orders of magnitude. Furthermore, as Solar Orbiter's orbital position and viewing angle continuously change, the characteristics of EUI data vary significantly, imposing strict adaptability requirements on compression algorithms.

\begin{figure*}
\centering
\includegraphics[width=1\textwidth]{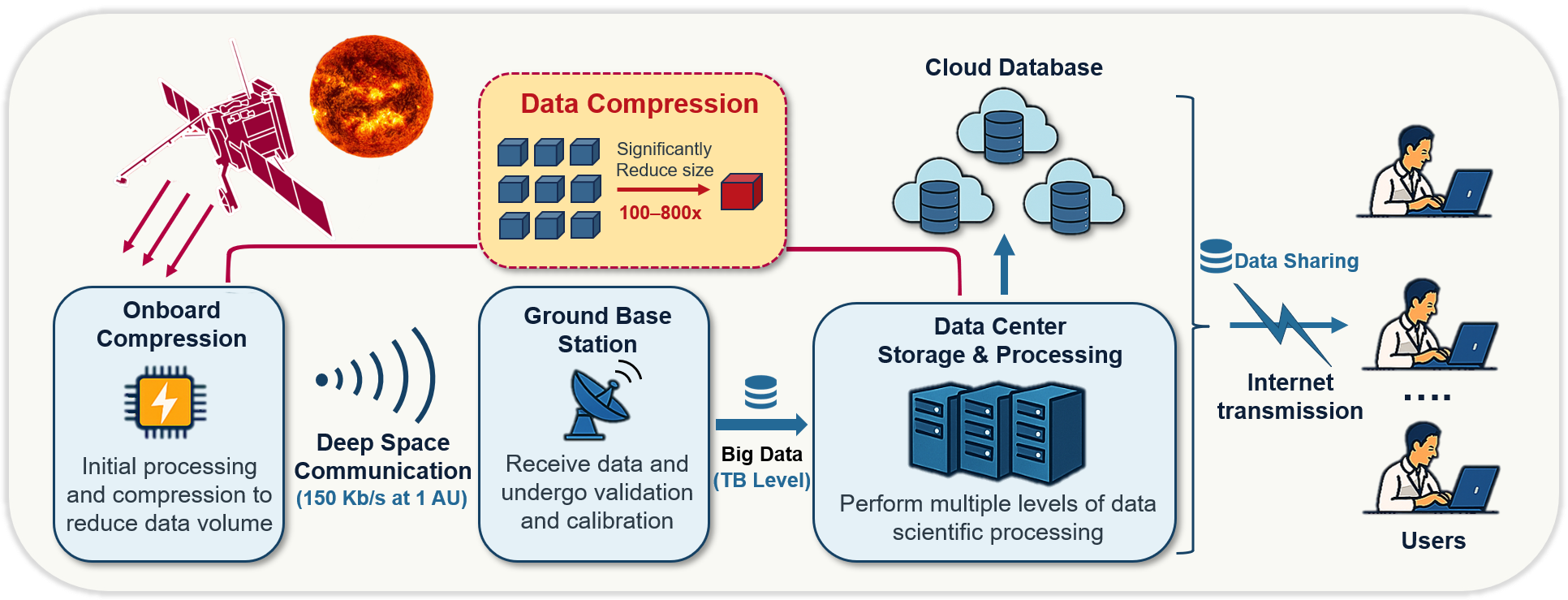}
\caption{ Life cycle of Solar Orbiter observation data. This figure illustrates the application of data compression techniques throughout the lifecycle of solar observation data. These techniques significantly reduce data volume, addressing communication and storage challenges, particularly for on-board systems and data centers.}
\label{fig1}
\end{figure*}

Compression techniques for scientific data can be broadly categorized into "lossless" and "lossy" approaches \citep{patel2015survey,di2024survey}. While lossless compression guarantees exact data reconstruction, it typically achieves limited compression ratios of only 2:1 to 3:1 for scientific floating-point data \citep{patel2015survey}. In contrast, error-bounded lossy compression can achieve much higher ratios, while maintaining scientific usability by controlling data distortion within user-specified tolerances \citep{cappello2019use,di2024survey}.

In recent years, a new generation of lossy compressors designed specifically for scientific data has emerged, including SZ \citep{di2016fast,tao2017significantly,liang2018error,zhao2021optimizing}, ZFP \citep{lindstrom2014fixed}, MGARD \citep{ainsworth2019multilevel}, and SPERR \citep{li2023lossy}. Unlike traditional lossy compressors such as JPEG \citep{wallace1991jpeg,taubman2002jpeg2000}, these error-bounded lossy compressors are designed to compress scientific data, while providing strict error control based on user requirements. These compressors have been successfully applied across various scientific domains. In climate simulation, \cite{baker2014methodology,baker2016evaluating,baker2017toward,baker2019evaluating} employed lossy compression on data produced by the Community Earth System Model. For cosmological simulations, \cite{pulido2019data,jin2020understanding,jin2021adaptive} proposed efficient schemes to reduce storage and transmission costs for Nyx and WarpX simulation codes. In astronomical observations, studies have evaluated how transform-based algorithms affect radio astronomy data quality \citep{peters2014impact,vohl2015data,chege2024impact}, while other researchers have explored efficient on-board compression algorithms for satellite missions using Cassini observational data \citep{xie2021compression, zhang2025high}.

Traditional compression methods have been explored for solar data. The RICE encoding algorithm \citep{rice1971adaptive}, which relies on basic preprocessing and encoding, achieves a maximum compression ratio of 20× for Solar Orbiter EUI data (refer to Section \ref{sec:3}). \cite{fischer2017jpeg2000} implemented JPEG2000, a wavelet-transform-based compression method, yet its compression ratio was limited to 30×, with significant quality degradation at higher compression levels. Recent efforts have explored machine learning approaches \citep{zafari2022attention,zafari2023neural,liu2024astrodllc,liu2024xception}, such as attention mechanisms and generative adversarial networks (GANs), achieving promising compression ratios but introducing substantial training and computational overhead.

Despite these advances, existing approaches for solar data compression have notable limitations: (1) they rely primarily on traditional lossy compression algorithms with relatively naive strategies, achieving limited compression ratios; (2) no comprehensive study has systematically applied or evaluated advanced error-bounded lossy compression techniques on solar EUV data; and (3) previous works lack interdisciplinary insights from both astronomy and computer science, failing to comprehensively illustrate the impact of lossy compression on scientific analyses of solar observations. To address these gaps, this paper introduces SolarZip, an efficient compression and evaluation framework for solar EUV imaging data. It features the following contributions:

\begin{itemize}
    \item We analyzed four advanced lossy compressors across two Solar Orbiter EUI datasets with 14 settings. These tools demonstrate clear advantages over traditional methods, such as RICE and JPEG2000, but still face limitations.
    
    \item We introduced SolarZip, a comprehensive compression and evaluation framework for solar EUV imaging data, particularly optimized for Solar Orbiter/EUI data.

    \item We designed a two-stage data evaluation framework that integrates strict error control with downstream scientific workflows: data distortion analysis and scientific post hoc analysis. This approach ensures that compressed data remain suitable for critical scientific research.
    
    \item We developed an adaptive hybrid compression strategy with optimized predictors to enhance compression quality. This method dynamically selects optimal compression methods based on different observational scenarios and user requirements, achieving a compression ratio of up to 800× reduction for FSI data and 500× for HRI$_{\text{EUV}}$ data.
    
\end{itemize}

This paper is structured as follows. Section~\ref{sec:2} introduces the fundamentals of data compression techniques. Section~\ref{sec:3} presents a comprehensive experimental comparison between advanced compression algorithms and traditional methods. In Section~\ref{sec:4}, we detail the SolarZip framework, including its compression workflow, hybrid strategy, and optimization methodology. Section~\ref{sec:5} provides a thorough evaluation of the SolarZip framework, demonstrating its superior performance through extensive experimental results. Finally, Section~\ref{sec:6} presents a conclusion to the paper and outlines promising directions for future research.

\section{Data compression foundation} 
\label{sec:2}
There are some traditional lossy compressors for images and videos, such as JPEG and MPEG, but they do not perform well in the context of scientific data. Error-bounded lossy compression represents a new generation designed for scientific data.
\subsection{Error-bounded lossy compression}
\label{sec:lossy}
Generally, error-bounded lossy compressors require users to set an error type, such as the point-wise absolute error bound and point-wise relative error bound, along with an error bound level (e.g., $\text{abs error}=10^{-3}$). The compressor ensures that the differences between the original data and the reconstructed data remain within the user-set error tolerance \citep{cappello2019use}. This ensures that the reconstructed data maintain a controlled level of accuracy, making it suitable for scenarios where precision is paramount.

The workflow of error-bounded lossy compressors can be summarized in the following steps: (1) data preprocessing, such as domain transformation and data blocking; (2) decorrelation via compression models, which are broadly categorized as either prediction-based or transform-based; (3) quantization with controlled error to achieve data compression; and (4) further lossless compression of quantized codes and other parameters using techniques such as arithmetic coding. The core component of a lossy compressor is its decorrelation model, as it critically influences both compression efficiency and speed.

The SZ family of compressors is a representative example of prediction-based compression models. Their predictors include linear regression predictors, the Lorenzo predictor, which was used in SZ1.4–SZ2.0 \citep{tao2017significantly,liang2018error}, and the spline interpolation predictor, introduced in SZ3 \citep{zhao2021optimizing,liang2022sz3}. Previous studies \citep{tao2019optimizing,liang2019significantly,zhao2020significantly} have explored the strengths and limitations of these predictors, as well as their cooperative use. Overall, the SZ series achieves high compression ratios, while maintaining a favorable compression speed, making it a versatile solution for scientific data compression.

Conversely, transform-based compression models employ different techniques to achieve decorrelation. For instance, ZFP \citep{lindstrom2014fixed} utilizes near-orthogonal transforms, while SPERR \citep{li2023lossy} applies the CDF 9/7 biorthogonal wavelet transform. ZFP is particularly notable for its high-speed performance; however, its limitation is that the compression ratio is constrained. The advantage of SPERR is that the hierarchical multidimensional DWT in SPERR can effectively capture the relevance between data points, which yields a high compression ratio after the SPECK encoding. One limitation of SPERR is that the transform and encoding processes have high computational costs and, hence, its execution speed is low, typically around 30\% of SZ3. Technical details related to compression can be found in Appendix~\ref{appendixb}.

\subsection{Applications of compression in solar observation data.}
Using Solar Orbiter as an example (Fig. \ref{fig1}), we explain how data compression techniques are integrated into the lifecycle of solar observational data. The process begins with data collection by the spacecraft’s remote-sensing and in situ instruments. Once generated, the data undergo initial processing and compression on-board. Since the space communication bandwidth is limited (nominal 150 kbit/s at 1 AU) and storage is constrained, compression is essential to facilitating efficient transmission. The data is downlinked via X-band telemetry to ESA’s deep-space ground stations. After reaching ground stations, they undergo initial validation and calibration. Subsequently, the data is transferred to dedicated scientific data centers, where scientists and researchers conduct various levels of data processing. Compression may also be applied at data centers to manage storage and transmission challenges posed by the vast volume of data. Finally, publicly available datasets are released through mission archives, enabling broader scientific access.

\begin{figure*}[htbp]
\centering

\includegraphics[width=\linewidth]{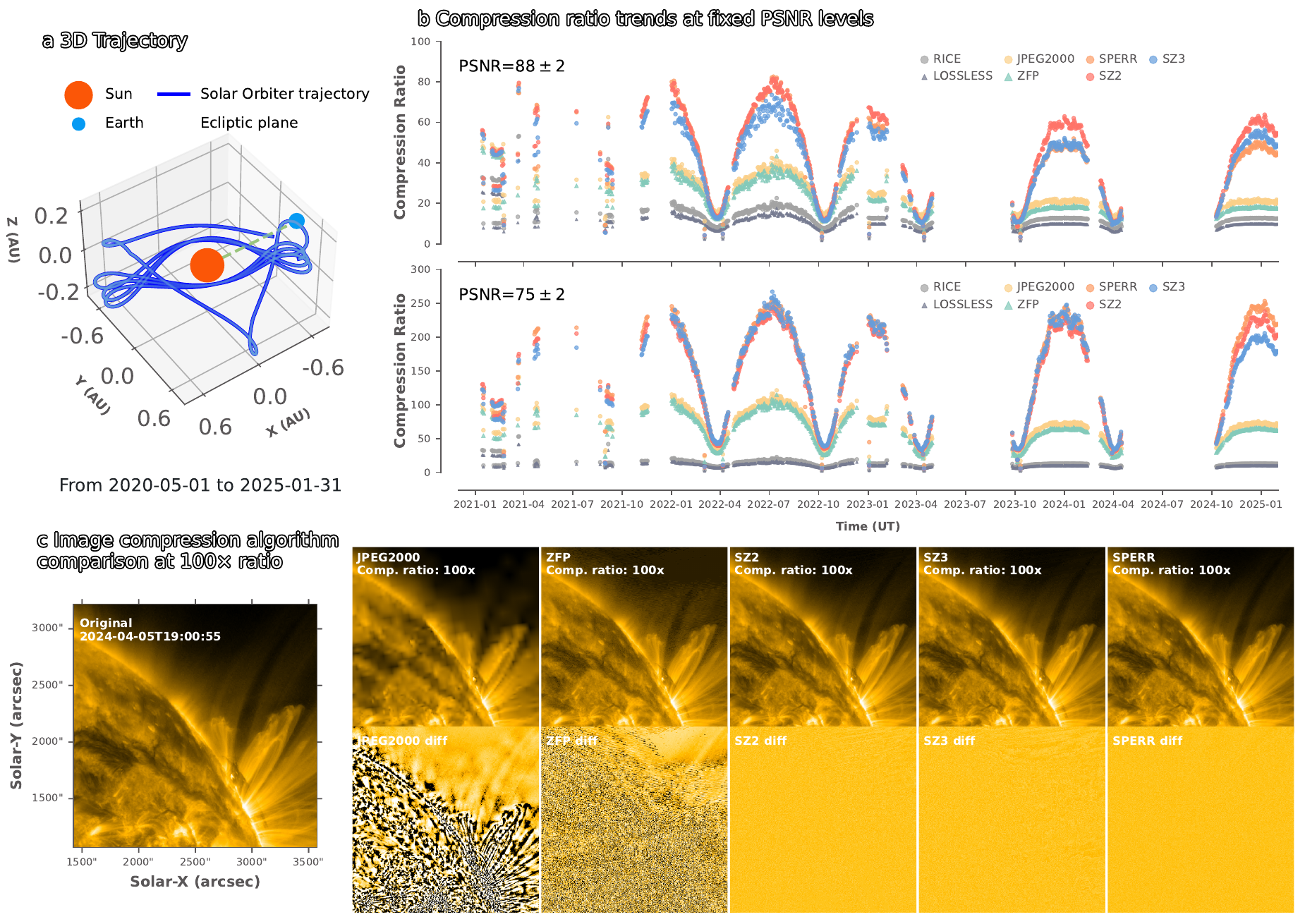}

\caption{
Panel a: Visualization of the Solar Orbiter’s trajectory based on the complete FSI dataset. Panel b: Compression ratio trends over time at a fixed quality level corresponding to a PSNR = 88 and PSNR = 75. Color coding indicates different compression methods. RICE and lossless compression (GZip) have fixed compression ratios and are used as baselines for reference. Advanced lossy compressors demonstrate significant advantages in compression ratio. Panel c: Visual comparison and difference maps between the original image (acquired on April 5, 2024) and the reconstructed images from five compression algorithms. All difference map display ranges were set to ±50 for consistency in this work.}

\label{fig2}
\end{figure*}

\subsection{Problem and metrics description}
\label{sec:metrics}
In this paper, we focus on the design and implementation of a lossy compression algorithm for solar EUV observational data represented by Solar Orbiter/EUI. The key goal is to achieve efficient data compression, significantly reducing the data volume. At the same time, it is crucial to ensure that the quality of the reconstructed data meets the requirements of scientific research. To achieve better compression methods, we employed the following metrics, which are widely used in prior literature and considered standard in the field \citep{leung1994review}.

Metric 1 \emph{CR}: We used the compression ratio (CR) to measure the compression performance. It is calculated by dividing the original data size by the compressed data size. A CR of 100 indicates that the compressed data is 1/100 the size of the original,\\
\begin{equation}
    \text{ }CR = \frac{\text{original size}}{\text{compressed size}}
    \label{cr}
.\end{equation}

Metric 2 \emph{PSNR}: For a data point, \(i\), we let \(e_{\text{abs}_i} = x_i - \tilde{x}_i\), where \([e_{\text{abs}}]\) is the absolute error. Also, we denote the range of \(X\) based on \(R_X\). To evaluate the average error in the compression, we first used the common root mean squared error (RMSE).\\
\vspace{-10pt}
\begin{equation}
      \text{ }RMSE = \sqrt{\frac{1}{N} \sum_{i=1}^{N} \left( e_{\text{abs}_i} \right)^2}
.\end{equation}
The peak signal-to-noise ratio (PSNR) is another commonly used average error metric for evaluating a lossy compression method \citep{berger2003rate}, especially in visualization. A higher value of the PSNR represents a lower error. It is calculated as\\
\vspace{-10pt}
\begin{equation}
\text{ }20 \cdot \log_{10}\left(\frac{R_X}{RMSE}\right).
\label{eq:psnr}
\end{equation}

Metric 3 \boldmath$\rho$: We employed the Pearson correlation coefficient, $\rho,$  to assess the linear correlation between the original and reconstructed datasets. A correlation coefficient of at least 0.9999 is generally required to ensure high fidelity,\\
\vspace{-12pt}
\begin{equation}
\text{ }\rho = \frac{\text{cov}(X, \tilde{X})}{\sigma_X \sigma_{\tilde{X}}}
.\end{equation}

Metric 4 \emph{SSIM}: The structural similarity index (SSIM) is another popular metric for evaluating the perceptual quality of images in compression. Unlike traditional error-based metrics, SSIM considers structural information, including luminance and contrast, which better reflects human visual perception. A higher SSIM value corresponds to greater similarity, with a value of 1 signifying perfect structural and perceptual equivalence. In this formula, $\mu_x$ and $\mu_y$ denote the mean intensities of images $x$ and $y$,  $\sigma_x^2$ and $\sigma_y^2$ are their variances, and $\sigma_{xy}$ is the covariance between them. The constants $C_1 = (K_1L)^2$ and $C_2 = (K_2L)^2$ are used to stabilize the division, where $L$ is the dynamic range of pixel values, and $K_1$, $K_2$ are small constants,

\begin{equation}
\text{ }SSIM(x, y) = \frac{(2 \mu_x \mu_y + C_1)(2 \sigma_{xy} + C_2)}{(\mu_x^2 + \mu_y^2 + C_1)(\sigma_x^2 + \sigma_y^2 + C_2)}
.\end{equation}

Post hoc metrics: In addition to conventional data distortion metrics,  this paper also introduces domain-specific scientific downstream analysis metrics. Aiming for a comprehensive evaluation of the impact of compression on EUI data, we provide a detailed discussion in Section \ref{sec4.4}.

\begin{figure*}
\centering
\includegraphics[width=0.9\textwidth]{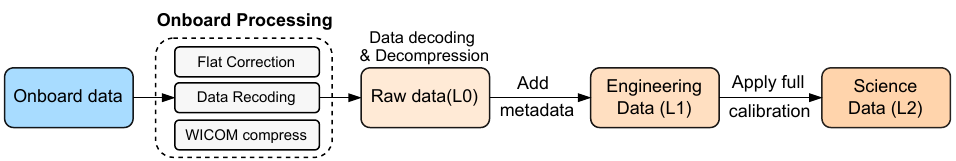}
\caption{Processing pipeline of EUI data (using FSI data as an example). The  on-board WICOM compression  (similar to JPEG2000) is mostly lossy, with only a small portion using lossless compression. Subsequent processing is performed on the ground.
}
\label{fig3}
\end{figure*}

\section{Comparison of advanced and traditional compression algorithms}
\label{sec:3}
In this section, we evaluate the performance of advanced lossy compression methods and traditional image compression techniques on EUI data. Based on the metrics specified in Section \ref{sec:metrics}, we perform a comparative analysis to assess the strengths and weaknesses of both classes of compressors.
\subsection{Experimental setup}
Environment: The advanced compression methods (error-bounded lossy compressors) included in the experiment are SZ2, SZ3, ZFP, and SPERR, while the traditional image compression techniques include JPEG2000 and RICE (native algorithm of the FITS format). As a reference, we also included the lossless compression algorithm GZip. The experimental environment is deployed on a cloud server equipped with two 16-core Intel Xeon Gold 6151 3.00GHz CPU, and 1,007 GB of RAM. Our test EUI dataset consists of two parts, described below.

FSI Dataset: EUI/FSI is an EUV imager observing the full solar disk in the 17.4 nm and 30.4 nm EUV passbands. It continually provides synoptic observations with a variable cadence depending on telemetry allocations and observing mode. We adopted a daily sampling strategy, selecting 17.4 nm images from the latest EUI data release 6.0 \citep{euidatarelease6}. During the data screening process, we excluded observations when FSI was operating in coronagraph mode to ensure data consistency and representativeness. The final FSI dataset totals 7.2 GB, covering observations from different orbital positions throughout Solar Orbiter's nearly 50 months from December 2021 to January 2025, comprehensively reflecting the characteristic variations in FSI data.

HRI$_{\text{EUV}}$ Dataset: The EUI/HRI$_{\text{EUV}}$ dataset is based on a flare observation campaign conducted on April 5, 2024. During this campaign, HRI$_{\text{EUV}}$ continuously observed a solar limb active region for approximately 4 hours from 19:59 to 23:59 with a temporal cadence of 16 seconds. This long time series observation generated a substantial amount of high-resolution data, totaling 4.2 GB. This dataset is particularly suitable for evaluating the performance of compression algorithms when processing complex, rapidly evolving solar active regions.

Before becoming publicly available, the raw EUI data undergo several preprocessing steps, which are illustrated in Fig.~\ref{fig3} \citep{Nicula2005, Poupat2013CWICOM}. Considering that the raw data are scarce and not publicly available, our main experiments are conducted on level-1 data, and all selected data have undergone the same on-board compression mode (lossy-high quality). To account for the potential impact of these prior processing steps, we also evaluated our method on level-1 data processed with different onboard compression mode in Appendix~\ref{appendixA}. Furthermore, in Appendix~\ref{appendixC}, we provide a mathematical justification to demonstrate the reliability of our compression results on the selected dataset.

\begin{figure*}
\centering
\includegraphics[width=0.9\textwidth]{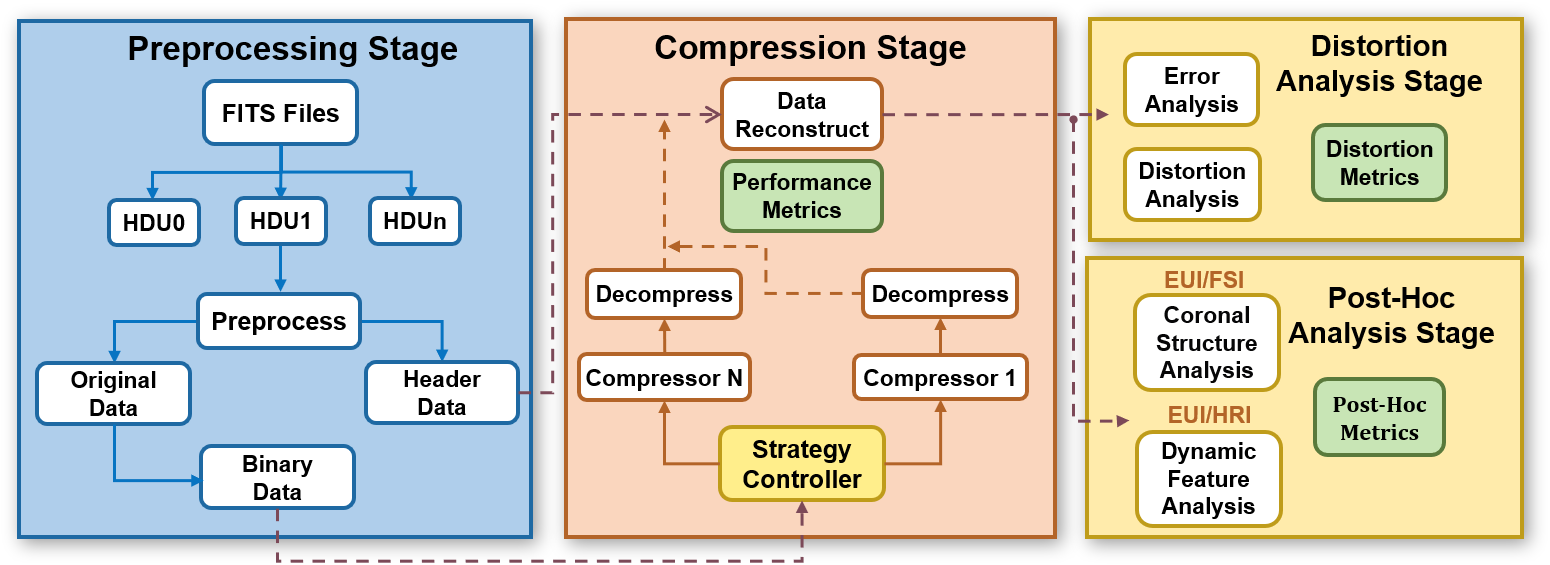}
\caption{Overview of SolarZip framework. The system consists of three stages: preprocessing, compression, and analysis. The core algorithmic innovation lies in the strategy controller, which can automatically tune and select the optimal compression strategy. The subsequent two stages of comprehensive analysis ensure that the decompressed data remain suitable for scientific purposes.
}
\label{fig4}
\end{figure*}

\subsection{Performance analysis and comparison}

To ensure a rigorous and objective comparison of compression performance across different algorithms, we adopted a unified quality metric PSNR (as Equation~\ref{eq:psnr}). A higher compression ratio at a given image quality indicates a superior algorithm. Figure \ref{fig2}b shows the compression ratios of different algorithms over observation time at a PSNR of 88. The SZ family and SPERR demonstrate the highest compression ratios, while JPEG2000 and ZFP achieve similar but inferior results compared to the best-performing algorithms. The RICE algorithm (native algorithm of the FITS format) exhibits the lowest compression ratio. Figure \ref{fig2}b shows the results at PSNR = 75, where the performance gap widens further. The best-performing algorithms achieve compression ratios close to 300×, whereas JPEG2000 is limited to 50×–100×. As an example, on January 1, 2023, SZ3 achieved a compression ratio 2.5× higher than JPEG2000 and 19.2× higher than RICE.

In Fig. \ref{fig2}c, we compare the reconstructed images of four advanced lossy compressors and JPEG2000 on the sample data. At the same compression ratio, JPEG2000 exhibits noticeable distortion and severe compression artifacts in many regions. In contrast, the four advanced compressors deliver superior performance. While ZFP shows minor deviations, the reconstructed images from SZ3, SZ2, and SPERR are visually indistinguishable from the original, maintaining excellent image quality.

\textit{Strength}: Experiments on full time series dataset and corresponding visual analyses confirm the superior performance of advanced compression algorithms on EUI data. Quantitative evaluation reveals that they significantly outperform traditional methods: achieving 10–30× higher compression ratios over RICE and 1.5–3× over JPEG2000, with equal reconstruction quality. This breakthrough is primarily attributed to the error-bound control mechanisms employed by these algorithms, which enable efficient data reduction while preserving image fidelity.

\subsection{Observation and motivation}
However, although advanced lossy compressors show clear advantages over traditional algorithms, we also observed their inherent limitations, which motivated the design of SolarZip.

\textit{Observation 1}: Compression ratios vary over time and closely follow changes in the spacecraft’s distance from the Sun. Specifically, when the spacecraft is farther from the Sun, the compression ratio increases under the same error bound; when it is closer, the compression ratio decreases. This can be explained by the reduced apparent size of the Sun in distant images, leading to a more uniform background distribution that is easier to compress.  

\textit{Observation 2}: No single compressor performs the best for EUI data. As shown in Fig. \ref{fig2}, at a PSNR of 75 dB, ZFP exhibits the lowest compression efficiency, while SZ3 and SPERR perform similarly well. At a higher PSNR threshold of 88 dB, the compression performance analysis reveals that SZ2 achieves superior results compared to other compressors, while SZ3 exhibits marginally lower performance than SPERR. We further investigated the reasons behind these variations. SZ3 relies on global interpolation and Lorenzo predictors, which perform well on datasets with strong global continuity. In contrast, SZ2 utilizes a block-based Lorenzo predictor, achieving higher accuracy for locally continuous data.

\textit{Motivation}: These observations confirm our key insight: EUI data characteristics are highly dependent on observation conditions, making it impossible to define a single optimal compression strategy. Therefore, a dynamic multi-compression framework is necessary to adapt to complex observational images and meet diverse scientific requirements.

\section{SolarZip compression framework}
\label{sec:4}
As discussed above, designing a comprehensive compression framework for  EUI data must address the diverse characteristic variations inherent in the data. This requires the dynamic selection of compression strategies tailored to varying data characteristics, while maintaining high compression efficiency. To this aim, we have proposed the SolarZip data compression and evaluation framework (Fig. \ref{fig4}).

\subsection{Overview of SolarZip}
The SolarZip workflow consists of four stages: (1) initialization, which is the pre-processing of FITS files and setting configurations; (2) compression, which consists of selecting the optimal compression strategy, automatically optimizing it,  then applying the compressor to the input data; (3) distortion analysis of the errors, evaluating distortion, and visualizing the results; and (4) post hoc analysis, performing downstream tasks such as a coronal structure analysis and dynamic feature analysis on reconstructed data.

Our workflow follows a modular design, enabling the efficient parallel compression of large volumes of FITS files. Users need to provide a configuration file and the FITS files. During the initialization stage, the system is configured based on the user’s configuration parameters and the FITS files are preprocessed. The preprocessing includes splitting the HDUs in the FITS files and extracting the data to be compressed. The data is then converted into binary format and passed to the next phase. FITS files are processed in batches, with the compression of each file occurring in parallel. Importantly, with the adaptive strategy controller, we are able to achieve the highest compression ratio for images from different observational scenarios and user requirements. Details on the adaptive strategy and optimization are presented in the next subsection.

After decompression, a distortion analysis is performed by comparing the decompressed data with the original, producing distortion metrics. The data processor then reconstructs the decompressed data back into FITS format, and the reconstructed FITS files are used for a further post hoc analysis.

\subsection{Adaptive hybrid compression strategy}
\label{subsec: ahcs}
We conducted a thorough evaluation of four lossy compressors on Solar Orbiter/EUI data (detailed in Section \ref{sec:3}). The results indicate that no single compressor consistently outperforms the others across various observational scenarios and scientific objectives. The core reason behind these variations lies in the distinct data prediction and transformation mechanisms of these advanced lossy compression algorithms (see more details in Subsection \ref{sec:lossy}).

In this subsection, we propose an adaptive optimization strategy that automatically selects the most suitable compression algorithm based on data characteristics in different scenarios. This strategy leverages a sampling-based approach to dynamically determine the optimal parameter configuration.

\begin{figure}
\centering
\includegraphics[width=9cm]{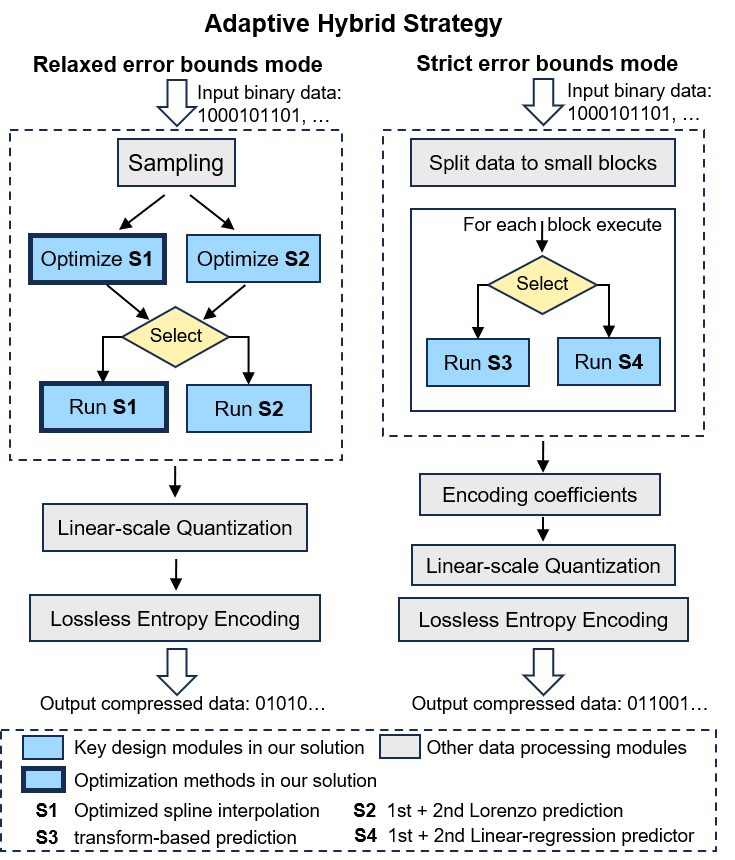}
\caption{Steps of the adaptive hybrid compression strategy. Left panel shows the compression strategy under relaxed error bounds, while the right panel shows strategy under strict error bounds. Our method dynamically selects the optimal compression strategy and optimize it. The different strategies are denoted by S1–S4 in the figure.
}
\label{fig5}
\end{figure}

Our adaptive hybrid compression strategy is depicted in Fig. \ref{fig5}. We categorize the precision requirements of EUI data into two types: relaxed and strict, using a relative error threshold of \(1 \times 10^{-4}\) as the boundary. For a relaxed error bound (\(eb > 1 \times 10^{-4}\)), the optimal strategy is chosen between spline interpolation prediction and Lorenzo prediction. Under a strict error bound (\(eb \leq 1 \times 10^{-4}\)), we select the best strategy from transform-based predictors and linear regression predictors. 

Specifically, we employed a heuristic sampling approach, where \(N\) data points are uniformly selected from the dataset to guide the compression strategy selection. For a relaxed error bound (\( \text{eb} > 1 \times 10^{-4} \)), the tuning process consists of four steps: (1) uniformly sampling \(N\) data points from the dataset; (2) optimizing the spline interpolation predictor (trial run S1 in Fig. \ref{fig5}) by selecting the best-fit interpolation method (linear or cubic) and optimizing the sequence of interpolation dimensions; (3) optimizing the Lorenzo predictor by dynamically selecting between first-order and second-order predictions; (4) selecting the best strategy with the highest compression ratio. For strict error bounds (\( \text{eb} \leq 1 \times 10^{-4} \)), the tuning process follows the same principle, except that run A corresponds to the transform-based predictor, while run B corresponds to the regression-based predictor.

Linear regression  and Lorenzo predictions have been effectively applied in previous studies and they are both block-based prediction methods. In this work, we conduct an offline analysis to determine the optimal block size of \(8 \times 8\), which is then set as the default parameter in our framework. Both linear regression and Lorenzo predictors support first-order and second-order prediction functions. Following the approach proposed in previous studies \citep{zhao2020significantly}, we dynamically select between first-order and second-order prediction functions for each data block. This ensures that every data block applies the optimal compression strategy. We applied the optimal orthogonal transformation in ZFP \citep{lindstrom2014fixed} as our prediction model because its de-correlation efficiency has been shown to be more effective than that of other transforms, such as the discrete cosine transform or wavelet transform.

\subsection{Optimization for spline interpolation predictor}
The compression method based on classical spline interpolation is able to achieve a high compression ratio under large error bounds. However, in some cases, noticeable image quality degradation occurs. For instance, in highly non-smooth regions of solar EUI images, such as flares, the interpolation predictor introduces visible compression artifacts in those areas. This issue arises because the basic interpolation-based predictor suffers from considerably low accuracy in long-range interpolation (\cite{liu2022dynamic}). Since it does not control the maximum stride length, the prediction accuracy becomes fairly low when the interpolation spans a long distance in the data array. To address these problems, SolarZip implements two key optimizations, which are described below. 

\textit{Grid-wise anchor points interpolation.} In the interpolation process, we specifically introduce grid-wise anchor points. Anchor points are predetermined data points, which are losslessly encoded and stored during compression. These anchor points divide the entire dataset into multiple blocks, and all other data points are predicted using points within a certain range, employing a multi-level interpolation method. This method effectively addresses the issues caused by long-range predictions. It is noteworthy that we found that if an appropriate stride is set for the anchor grid, the overhead associated with storing the losslessly compressed anchor points becomes negligible. More details on the anchor points interpolation are described in Appendix ~\ref{appendixb3}.

\textit{Level-wise interpolation with error bound auto-tuning.}
We set different error bounds at different levels of interpolation (as opposed to the unified error bounds used in SZ3). In our two- (2D) data, 75\% of the data points fall within the lowest interpolation level (level 1), which are predicted by higher-level reconstructed data points, while the remaining 25\% of the data points are predicted at higher levels. Therefore, setting smaller error bounds at higher levels helps ensure the overall prediction accuracy, thereby improving the compression quality, 
\begin{equation}
    \text{ }e_l = \frac{e}{ min(\alpha^{l-1}, \beta)} \quad (\alpha \geq 1 \text{ and } \beta \geq 1)
    \label{leavel error}
.\end{equation}
The level-wise error bounds \(e_l\) are dynamically adjusted based on Equation \ref{leavel error}. The parameters \(\alpha\) and \(\beta\) are introduced, where \(e\) represents the global error bound set by the user. We perform offline testing with parameter sets \(\alpha = \{1, 1.5, 2\}\) and \(\beta = \{2, 3, 4\}\), comparing the bit-rate and PSNR values across different parameter configurations. Ultimately, we select \(\alpha = 1.5\) and \(\beta = 4\) as our optimal parameters.

\subsection{Post hoc analysis stage}
\label{sec4.4}
In addition to the standard distortion analysis, we implemented specialized post hoc analysis tailored to the scientific requirements of solar physics research.

\begin{table*}[h!]
\centering
\caption{Comparison of four compressors on FSI and HRI$_{\text{EUV}}$ datasets under three error bounds. }
\footnotesize
\begin{tabular*}{0.95\textwidth}{l l c c c c c c c c}
\toprule
\multirow{2}{*}{\textbf{Error-bound}} & \multirow{2}{*}{\textbf{Compressor}} & \multicolumn{4}{c}{\textbf{FSI dataset}} & \multicolumn{4}{c}{\textbf{HRI$_{\text{EUV}}$ dataset}} \\ 
\cmidrule(lr){3-6} \cmidrule(lr){7-10}
& & \textbf{Comp ratio} & \textbf{PSNR} & \textbf{Coefficient} & \textbf{SSIM} & \textbf{Comp ratio} & \textbf{PSNR} & \textbf{Coefficient} & \textbf{SSIM} \\ 
\midrule
\multirow{5}{*}{$1e-4$} 
& SPERR & 44.36 & 91.74 & 1.00000 & 0.74 & 5.69 & 90.78 & 1.00000 & 0.99 \\ 
& ZFP & 33.21 & 100.84 & 1.00000 & 0.84 & \textbf{5.83} & 103.2 & 1.00000 & 1.00 \\ 
& SZ3 & 42.47 & 89.29 & 1.00000 & 0.76 & 5.14 & 85.33 & 1.00000 & 0.99 \\ 
& SZ2 & 45.86 & 90.07 & 1.00000 & 0.77 & 4.83 & 84.79 & 1.00000 & 1.00 \\ 
\rowcolor{myblue!18}
&  \textit{\textbf{SolarZip}} &  \textbf{57.62} &  92.88 &  1.00000 &  0.76 &  5.71 &  90.14 &  1.00000 &  0.99 \\ 
\midrule
\multirow{5}{*}{$1e-3$} 
& SPERR & 151.98 & 77.23 & 0.99999 & 0.45 & 10.51 & 65.57 & 0.99997 & 0.96 \\ 
& ZFP & 63.42 & 82.33 & 0.99999 & 0.58 & 10.75 & 73.08 & 0.99999 & 0.99 \\ 
& SZ3 & 149.55 & 73.99 & 0.99998 & 0.46 & 10.40 & 65.04 & 0.99997 & 0.96 \\ 
& SZ2 & 142.32 & 74.91 & 0.99998 & 0.56 & 10.28 & 64.79 & 0.99996 & 0.96 \\ 
\rowcolor{myblue!18}
&  \textit{\textbf{SolarZip}} &  \textbf{155.82} &  76.54 &  0.99999 &  0.46 &  \textbf{11.04} &  69.75 &  0.99998 &  0.97 \\ 
\midrule
\multirow{5}{*}{$1e-2$} 
& SPERR & 660.23 & 61.72 & 0.99981 & 0.39 & 112.53 & 49.98 & 0.99928 & 0.39 \\ 
& ZFP & 221.82 & 68.20 & 0.99987 & 0.46 & 16.17 & 61.12 & 0.99993 & 0.816 \\ 
& SZ3 & 685.96 & 57.17 & 0.99952 & 0.42 & 79.56 & 48.37 & 0.99869 & 0.38 \\ 
& SZ2 & 560.12 & 57.87 & 0.99947 & 0.39 & 82.26 & 46.08 & 0.99897 & 0.35 \\ 
\rowcolor{myblue!18}
&  \textit{\textbf{SolarZip}} &  \textbf{698.34} &  60.92 &  0.99984 &  0.41 &  \textbf{115.78} &  57.80 &  0.99974 &  0.34 \\ 
\bottomrule
\label{tab1}
\end{tabular*}
\tablefoot{
 Key metrics include compression ratio (higher=better), PSNR (higher=better), correlation coefficient (1.0=perfect), and SSIM (1.0=perfect).
}
\end{table*}

\subsubsection{FSI large-scale coronal structure analysis}
For the FSI data, we focus on evaluating how compression affects large-scale dynamic structures in the solar corona. We implement a circular intensity extraction method, where a virtual circle is placed at 1.05 solar radii from the disk center, corresponding to the lower corona region where many important dynamic phenomena occur. Intensity values are sampled along this circle to generate a 1D intensity profile that captures the coronal structures.

The intensity profiles from the original and reconstructed compressed images are then compared to assess how different compression algorithms preserve the coronal features. We compare several metrics, including:
\begin{itemize}
\item morphology-based visual comparison of full-disk FSI images;
\item intensity profile correlation coefficient.
\end{itemize}
This analysis is particularly important for studying large-scale coronal evolution and identifying the onset of coronal mass ejections, where subtle changes in intensity distribution can have significant scientific implications.

\subsubsection{HRI$_{\text{EUV}}$ small-scale dynamic feature analysis}

For the HRI$_{\text{EUV}}$ data, our post hoc analysis focuses on the preservation of small-scale dynamic structures in selected frames. We examine features such as plasma flows and fine magnetic structures, which are crucial for understanding energy transport and release in the solar atmosphere.
We implement feature tracking algorithms to identify and characterize dynamic features in both the original and compressed reconstructed data. The analysis includes:
\begin{itemize}
\item morphology-based visual comparison of local dynamic features in HRI$_{\text{EUV}}$ images;
\item difference maps between original and compressed images;
\item median and standard deviation of the pixel difference distribution in the difference maps.
\end{itemize}
In contrast to FSI's analysis of large-scale structures, for HRI$_{\text{EUV}}$, we focus on a detailed analysis of representative small-scale structures, comparing the sensitivity of different features under various compression ratios to propose targeted dynamic compression strategies.

\section{Evaluation results}
\label{sec:5}
We tested the compression performance and reconstruction quality of SolarZip alongside five advanced lossy compression techniques on EUI data. Our evaluation framework consists of two analytical stages. The first stage calculates compression performance metrics by comparing the original and reconstructed data, while the second stage incorporates expert domain knowledge for systematic post hoc analysis, ensuring a comprehensive multidimensional assessment of the compression performance. Additionally, we conducted simulation experiments based on the hardware conditions of Solar Orbiter/EUI, verifying the significant improvement in data transmission efficiency enabled by lossy compression algorithms.

This detailed examination evaluates whether the compressed data retains sufficient information to support the scientific analysis of transient solar phenomena and small-scale structures, which is essential for studies of magnetic reconnection, wave dynamics, and plasma heating mechanisms in the solar atmosphere. By combining these specialized post hoc metrics with standard distortion analysis, we provide a comprehensive evaluation of compression performance that directly addresses the scientific use cases for EUI data.

\subsection{Evaluation of compression performance}

Table \ref{tab1} shows the test results of SolarZip against four error-bounded lossy compression algorithms (ZFP, SZ3, SZ2, and SPERR) on the FSI dataset and HRI$_{\text{EUV}}$ dataset. The experiments show that SolarZip achieves optimal compression ratios on both datasets, with an improvement of 5.6-30.4\% over the second-best compressor, while maintaining excellent fidelity (PSNR > 60 dB). This advantage is mainly due to our proposed adaptive hybrid compression strategy (AHCS), which demonstrates excellent adaptability to the data and can optimize and select the optimal compression strategy, as well as the optimization of the spline interpolation predictor, which improves the quality of the reconstructed images under high error bounds. Thus, SolarZip achieves a high compression ratio for solar science data while ensuring scientific usability.

\begin{figure*}[htbp]
\centering
\includegraphics[width=1\textwidth]{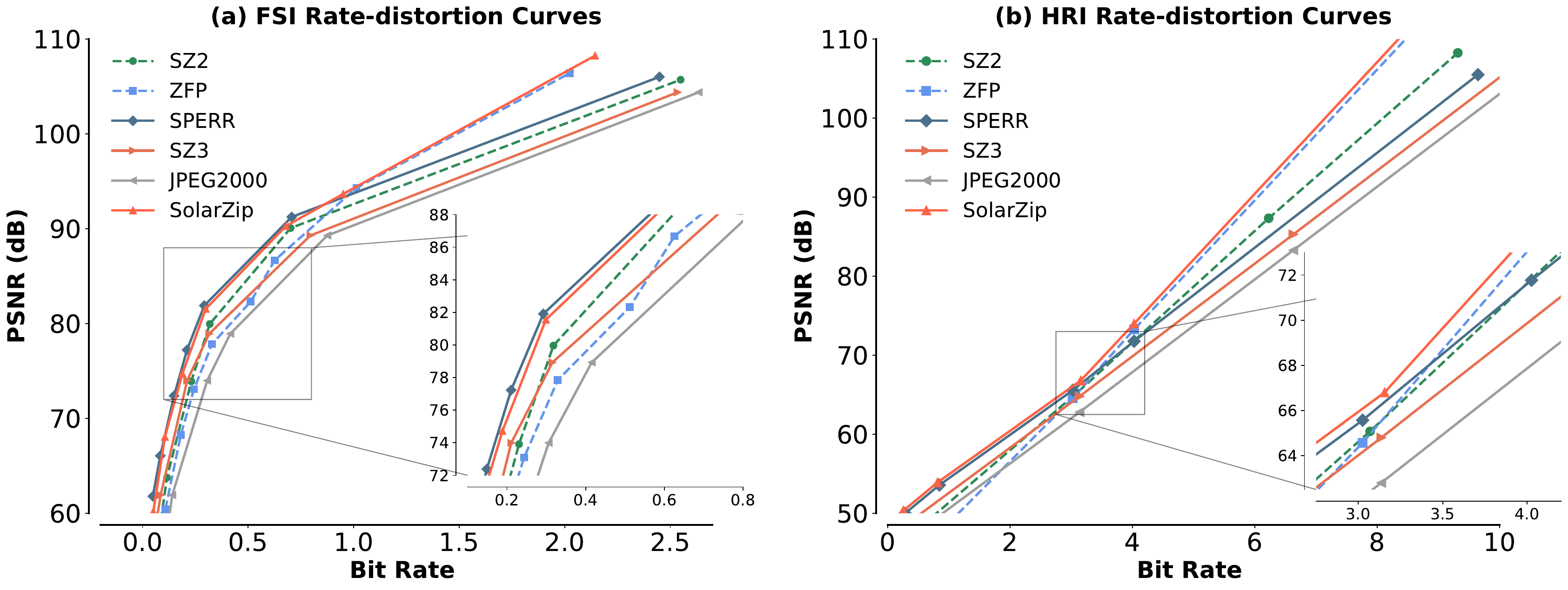}
\caption{Rate-distortion curves on different datasets. Different compressors are distinguished by color, with our method indicated by the red line. A higher PSNR corresponds to better image quality at the same bit rate. SolarZip demonstrates the best overall performance on both datasets.}
\label{fig6}
\end{figure*}

To enhance the rigor and objectivity of the compression performance comparison, we plot the rate-distortion curves for our solution and other lossy compressors, comparing the distortion quality at the same rate. Here, rate refers to the bit rate in bits per value, and we use the PSNR to measure the distortion quality. The PSNR is calculated by Equation~\ref{eq:psnr} in decibels. Generally speaking, in a rate-distortion curve \citep{berger2003rate}, a higher bit rate indicates that more bits are required to store each value, resulting in higher quality of the reconstructed data after decompression, as reflected by a higher PSNR.

As discussed in Section~\ref{subsec: ahcs}, we designed an adaptive hybrid strategy to optimize the compression quality across the entire bit-rate range. Fig.~\ref{fig6}a presents the rate-distortion curves of our algorithm compared to five other methods on the FSI dataset. The results demonstrate that our adaptive hybrid strategy plays a crucial role in improving compression quality. As shown in Fig.~\ref{fig6}  a, our compression algorithm achieves near-optimal quality across almost all bit rates. Particularly, for bit rates below 1.0, our method exhibits notably superior compression quality compared to SZ3 and SZ2, attributed to its dynamic selection between the automatically optimized spline interpolation predictor and the Lorenzo predictor. Moreover, our approach performs comparably well to the SPERR compressor, which is specifically designed for high-quality compression but is unsuitable for our task due to its inherent limitations, as discussed later in this work. At a PSNR of 77, our method achieves a compression ratio 300\% higher than JPEG2000, demonstrating the efficacy of our spline interpolation predictor optimizations. When the bit rate exceeds 1.0, our solution surpasses SPERR, becoming the most rate-distortion efficient compressor. This is because, under a relaxed error bound greater than $1 \times 10^{-4}$, our strategy accurately selects the linear regression predictor and the transformation-based predictor (using the same orthogonal transformation matrix as ZFP). Notably, our method consistently outperforms JPEG2000, a widely used compression standard in astronomy.

Fig.~\ref{fig6}b presents the rate-distortion curves of all compressors on the HRI$_{\text{EUV}}$ dataset. The results demonstrate that our solution achieves the best performance among all six compressors. It can be observed that at a bit rate of approximately 3, our rate-distortion curve exhibits a distinct inflection point, while maintaining superior quality. This behavior results from an adaptive adjustment in our selection strategy as the error bounds transition from a strict to a relaxed mode, with the threshold determined through offline analysis. Additionally, at the same PSNR level (e.g.,  equal to 60), our method achieves a 50\% higher compression ratio than JPEG2000, further validating the effectiveness of our optimizations on the spline interpolation predictor in improving the image quality.

\begin{figure}[htbp]
\centering
    \centering
    \includegraphics[width=\linewidth]{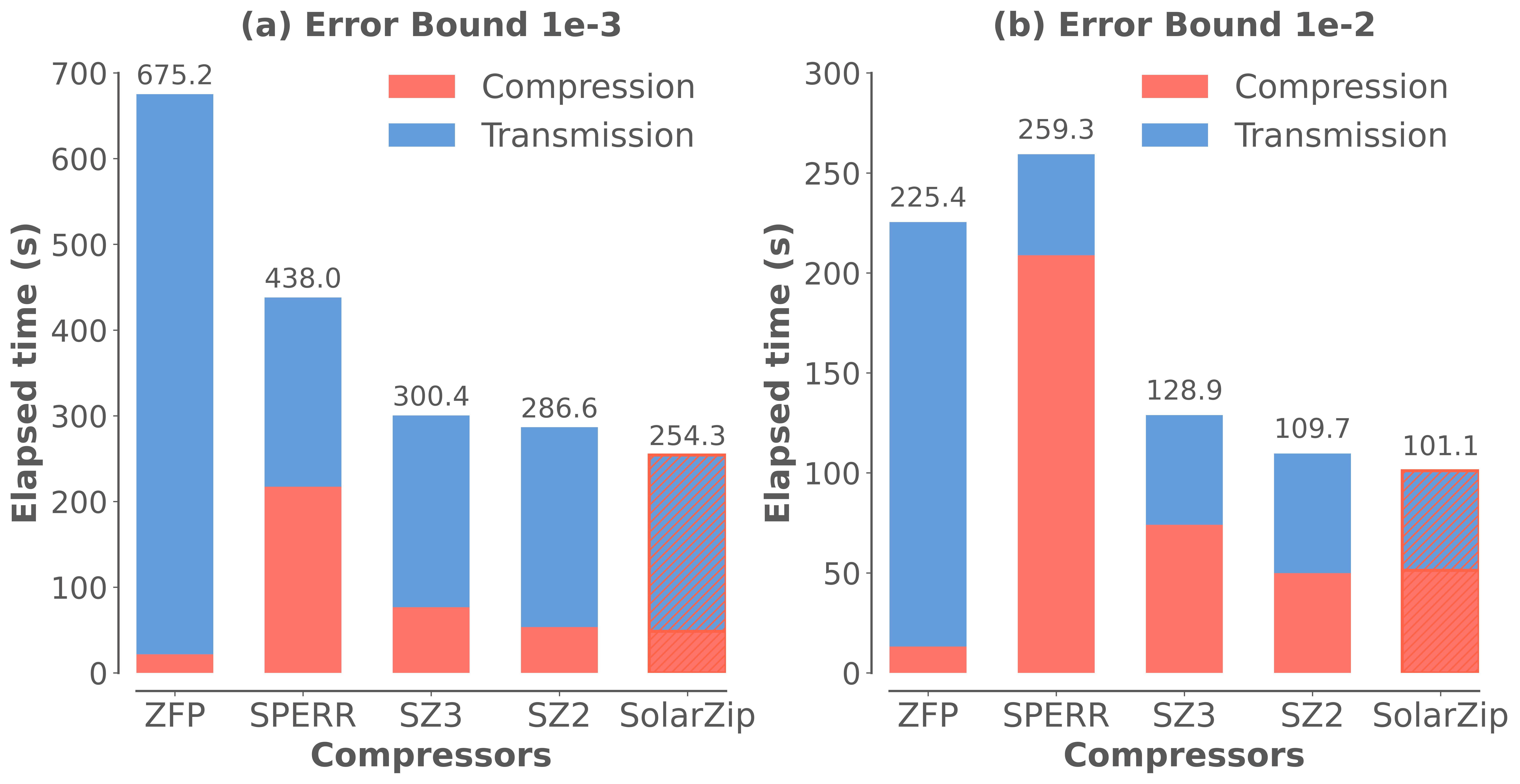}
\caption{Comparison of elapsed transmission times for different compressor with a data volume of 1000 MB. The elapsed time includes compression time (red) and transmission time (blue), while SolarZip demonstrates the highest efficiency, reducing the total time by 270×.}
\label{fig7}
\end{figure}

\begin{figure*}
\centering
\includegraphics[width=1\textwidth]{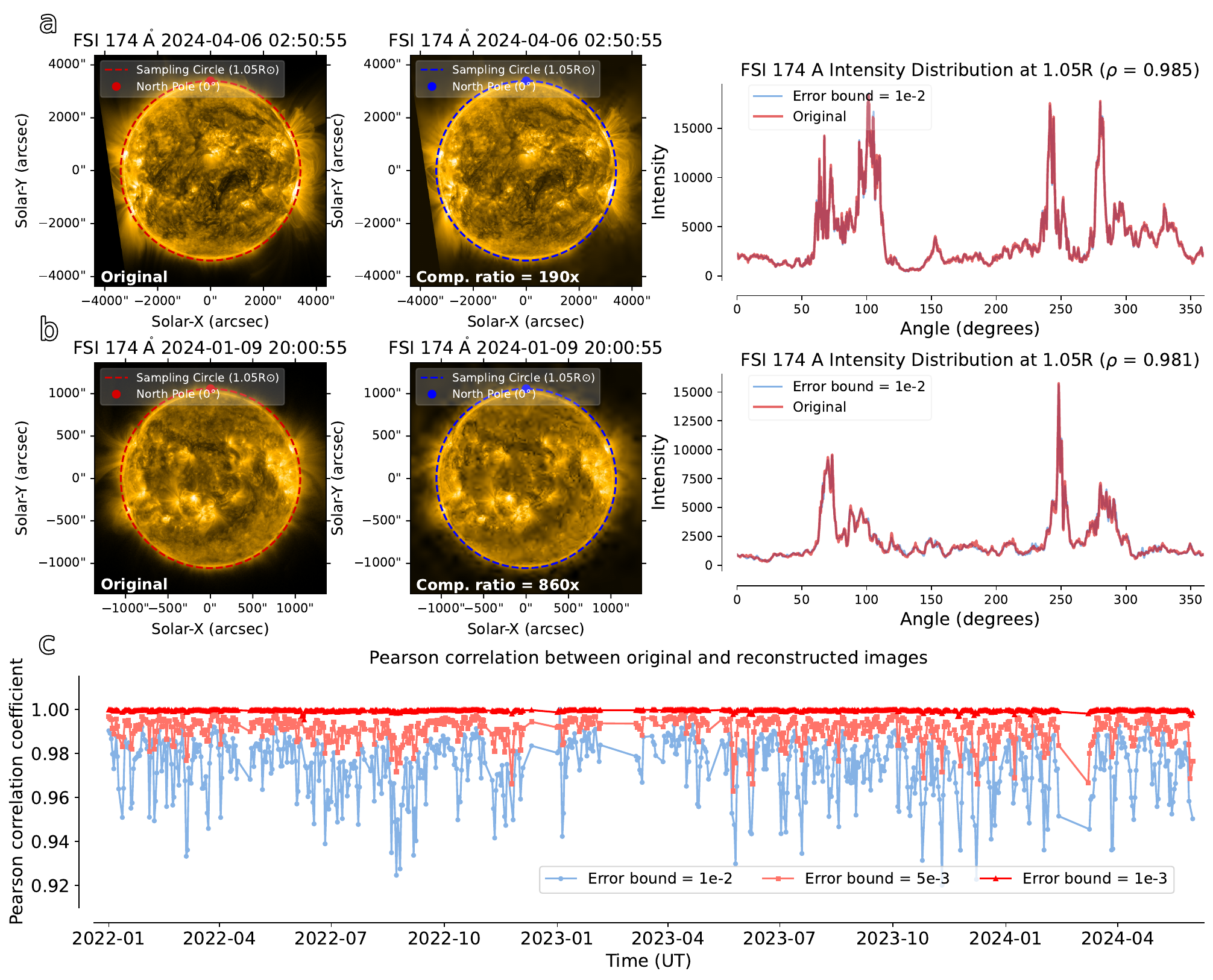}
\caption{Post hoc analysis comparison results of FSI. Panel a: Comparison image near perihelion on April 6, 2024, with red and blue dashed circles extracting intensity distributions at 1.05 solar radii from the original and reconstructed images. Results are shown in the right panel. Panel b: Comparison image near aphelion on January 9, 2024, with red and blue dashed circles extracting intensity distributions at 1.05 solar radii from the original and compressed images. Results are displayed in the right panel. Panel c: Correlation coefficients of intensity distributions at 1.05 solar radii between original and reconstructed images under three different error bounds, based on 2.5 years of FSI data.}
\label{fig8}
\end{figure*}

We conducted a simulation experiment on the compression and data transmission process aboard the Solar Orbiter. Solar Orbiter/EUI is powered by a WICOM compression ASIC with SDRAM, delivering an average downlink telemetry rate of approximately 300 Kbit/s to ground stations \citep{rochus2020solar, marirrodriga2021solar}. We simulated the data transmission process after compression and recorded the total time, including both compression and transmission durations. In Fig.~\ref{fig7}, we present the results under two different error bounds. Our solution achieved the shortest elapsed time, demonstrating the highest data transmission efficiency. For instance, transferring 1000 MB of uncompressed data takes approximately 7.5 hours, whereas our method completes the task in just 101 seconds, improving the efficiency by a factor of 270. This is due to the superior compression performance and high compression speed of our method. Although SPERR achieved a slightly higher compression ratio than our solution, its slow compression speed resulted in a significantly longer total time.

\subsection{Evaluation of the post hoc analysis}
After evaluating generic compression performance, our focus shifts to whether reconstructed FSI and HRI$_{\text{EUV}}$ images can meet the requirements of solar physics observational research. Based on the data's inherent variations and specific research content, we propose what we consider appropriate dynamic compression strategies.

\begin{figure*}[!htbp]
\centering
\includegraphics[width=1\textwidth]{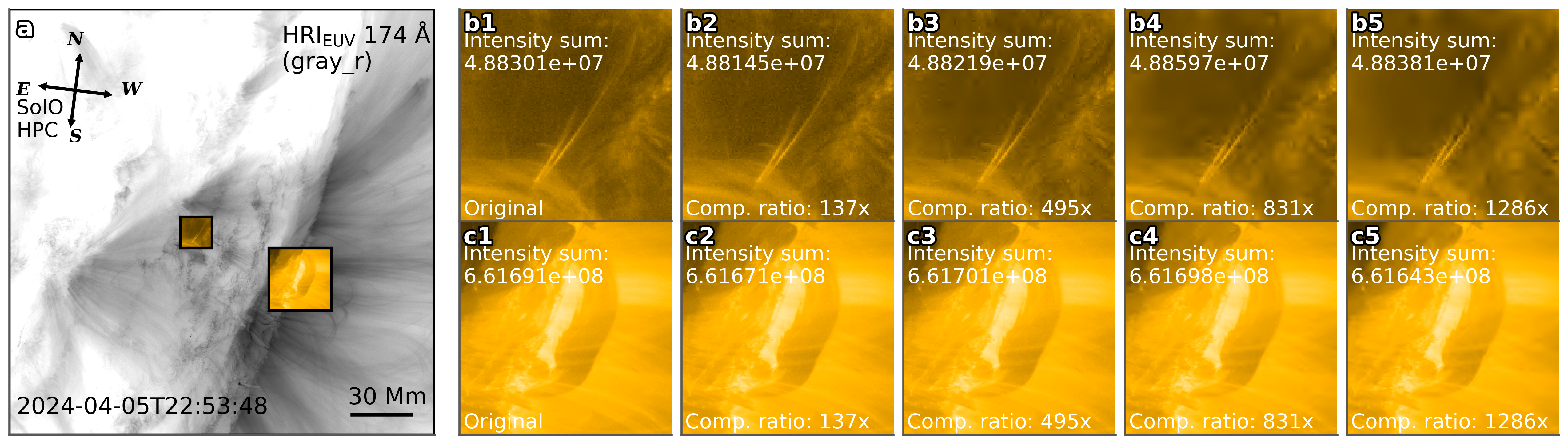}
\caption{Post hoc analysis comparison results of HRI$_{\text{EUV}}$. Panel a:  Selected demonstration image showing two representative features: a jet 
(area of 150 square pixels) and a prominence (area of 300 square pixels). Panel b: Comparison between the original image and images at four different compression ratios in the jet region, with annotations showing compression ratios and the sum of pixel intensities within the region. Panel c: Comparison between the original image and images at four different compression ratios in the prominence region, with annotations showing compression ratios and the sum of pixel intensities within the region.}
\label{fig9}
\end{figure*}

\begin{figure*}[!htbp]
\centering
\includegraphics[width=0.95\textwidth]{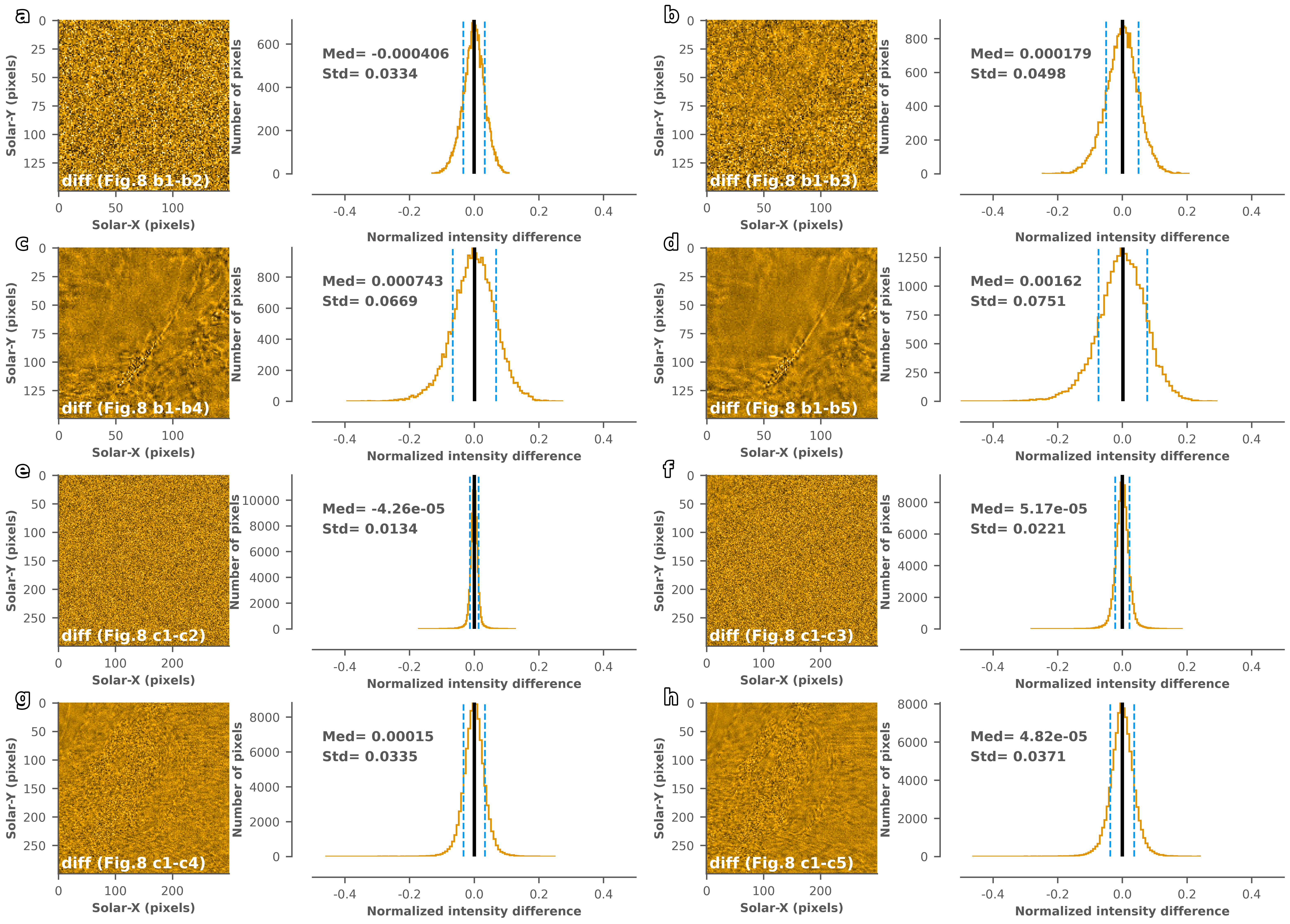}
\caption{Panels a-h: Difference maps between original and compressed images, labeled with corresponding panels from Fig. \ref{fig9}. Right side of difference maps: Pixel difference distribution of the difference maps, annotated with median and standard deviation. Black solid line indicates the median position, blue dashed lines represent the range of one standard deviation on either side of the median.}
\label{fig10}
\end{figure*}

\subsubsection{FSI reconstructed image analysis}
As shown in Fig. \ref{fig8}, we present two sets of FSI comparison images from April 6, 2024 (near perihelion) and January 9, 2024 (near aphelion), with the same error bound of 1e-2. We find that at high compression ratios (190× and 860×), no significant differences are visible in the perihelion image (Fig. \ref{fig8}a), while the aphelion  reconstructed image shows discrepancies in coronal morphology (appearing as discontinuities in the coronal structure, Fig. \ref{fig8}b). This occurs because at perihelion, the closer distance provides  a resolution for the solar disk that is approximately three time higher compared to aphelion, resulting in lower compression ratios near perihelion. We extracted the intensity curves from a circle at 1.05 solar radii in the FSI images and compared the correlation coefficients between original and compressed image intensity curves (shown in the right panel). We found that despite the apparent morphological information loss in the aphelion images (with higher compression ratios), they still maintain high correlation coefficients with the original image intensity curves. In Fig. \ref{fig8}c, we present the pearson correlation coefficients between the original and reconstructed intensity distributions at 1.05 solar radii over the complete 30-month FSI dataset under three error bounds. At an error bound of 1e-3, the correlation remains extremely close to 1 (average of 0.99998). Even under a looser bound of 1e-2, although the coefficient is lower, our visual inspection confirms that the resulting reconstruction errors still remain within acceptable levels.

\begin{figure}
\centering
\includegraphics[width=0.45\textwidth]{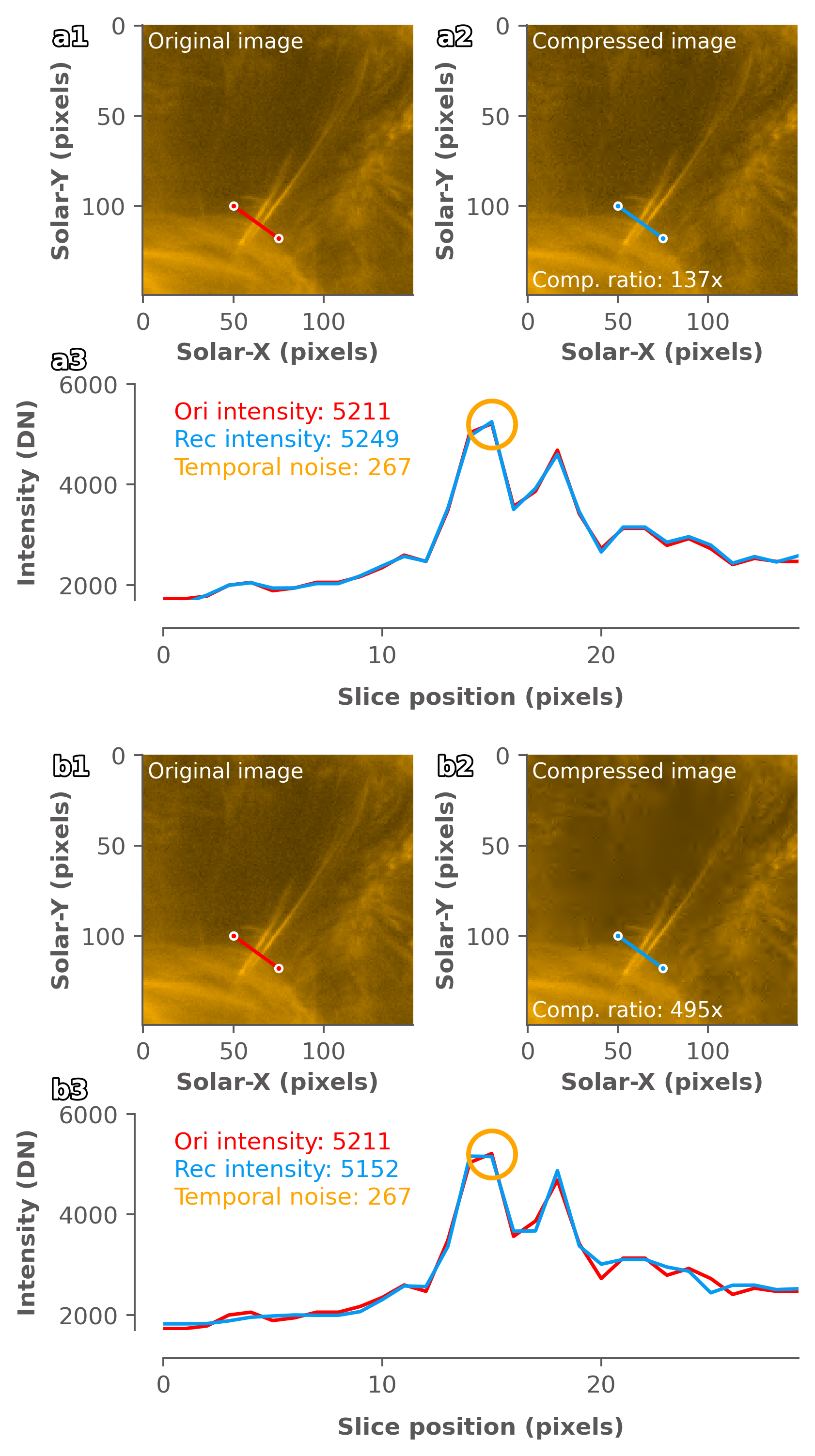}
\caption{Comparison of algorithm-induced errors versus HRI$_{\text{EUV}}$ temporal noise errors. Panel a: Intensity distributions extracted across the jet structure from both the original HRI$_{\text{EUV}}$ image and the image with 137× compression ratio. Panel b: Intensity distributions extracted across the jet structure from both the original image and the image with 495× compression ratio. The intensity maxima in the cross-jet intensity distributions are marked for each image.}
\label{fig11}
\end{figure}

Unlike traditional coronal EUV imagers, FSI has an unprecedentedly large FOV: $(228')^2$ \citep{rochus2020solar,berghmans2023first}, which has significant overlap with the Solar Orbiter coronagraph Metis \citep{antonucci2020metis}. At perihelion, this FOV corresponds to $(4\,R_{\odot})^2$ such that the full solar disk is always seen, even at maximum off-pointing $(1\,R_{\odot})$. This FOV is significantly wider than the $(3.34\,R_{\odot})^2$ of EUVI \citep{howard2008sun} or the $(3.38\,R_{\odot})^2$ of SWAP \citep{seaton2013swap}. When close to aphelion, this FOV corresponds to $(14.3\,R_{\odot})^2$, providing unique opportunities to image the middle corona and eruptions that transit through this region.

Our post-analysis study demonstrates that for FSI with such dynamically varying field of view and observational targets, our SolarZip algorithm achieves compression ratios of nearly 200× for perihelion FSI images at the error bound of 1e-2, while aphelion FSI images reach ultra-high compression ratios exceeding 800×. Considering the dynamic variations of FSI data, users can lower the error bound when Solar Orbiter approaches the Sun to achieve higher fidelity, while increaseing the error bound when it moves away from the Sun to reduce fidelity and maintaining higher compression ratios, keeping the overall compression performance within an optimal range.

\subsubsection{HRI$_{\text{EUV}}$ reconstructed image analysis}
The HRI$_{\text{EUV}}$ plate scale is $0.492''$, which implies unprecedented ultra-high resolution EUV observations. On April 5, 2024, Solar Orbiter reached a distance of 0.29 AU from the Sun, giving (single) pixel values on the Sun of $(105\,\text{km})^2$ for HRI$_{\text{EUV}}$. This unparalleled resolution provided us with an opportunity to study small-scale dynamic structures in the solar corona. We selected the frame at 22:53:48 UT as a representative image to analyze whether the reconstructed image meets the actual scientific requirements (Fig.~\ref{fig9} a). We selected two regions for comprehensive analysis, covering two well-studied phenomena widely present in the solar atmosphere: solar jets and prominences.

Solar jets, defined as collimated, beam-like plasma ejections along magnetic field lines, are ubiquitous in all regions of the solar atmosphere, including active regions, coronal holes, and quiet-Sun regions \citep{2016SSRv..201....1R,2021RSPSA.47700217S,2021PhDT........23J,2022MNRAS.516L..12T,2023MNRAS.520.3080T,2024ApJ...963....4S}. Prominences, as widely present magnetized plasma structures in the solar atmosphere, exhibit rich dynamic characteristics \citep{2011LRSP....8....1C,2012LRSP....9....3W,2015LRSP...12....3W,2020ChSBu..65.3909S,2012ApJ...745L..18A}. We compared the reconstructed and original images of the jet region and prominence region under different error bound values (corresponding to different compression ratios; Fig. \ref{fig10}b and c). First, we can find that the intensity and of both the original image and the reconstructed image at different compression rates remain highly consistent. For the smaller jet region (150 square pixels), we observed noticeable blurring effects beginning at a compression ratio of 831×, while both 137× and 495× compression ratios maintained morphological consistency. For the larger prominence region (300 square pixels), our morphology-based visual inspection shows that the reconstructed images maintained excellent prominence structural features compared to the original image, even at the highest compression ratio of 1286×.

To analyze the differences between original and reconstructed HRI$_{\text{EUV}}$ images, we employed two complementary methods. First, we generated difference maps by subtracting the reconstructed images from the originals, highlighting structural discrepancies in selected regions containing solar jets (Fig. \ref{fig10}a-d, left panels) and prominences (Fig. \ref{fig10}e-h, left panels). This enabled direct morphological comparison of fine-scale features. Second, we computed the normalized intensity difference (1-(original and reconstructed)) and plotted its histogram (Fig. \ref{fig10}, right panels). We computed the median and standard deviation to characterize the compression artifacts, with vertical lines marking the median and $\pm 1\sigma$ boundaries. The two structures exhibited different sensitivities to different compression ratios. At compression ratios of 137× and 495×, difference images of the jet region exhibit no distinct structural features, appearing primarily as random noise. However, at 1286×, elongated jet structures emerged in the difference map (Fig. \ref{fig10}d), indicating significant discrepancies at the jet edge. In contrast, the prominence region difference images only begin to show subtle prominence structures at the 1286× (Fig. \ref{fig10}h); yet the histogram at this compression ratio maintains a relatively narrow distribution, supporting our visual assessment of preserved morphological features.

In our analysis of the jet region, we found that the difference image at a compression ratio of 137 showed almost no jet structures (Fig. \ref{fig11}a), instead exhibiting features of noise signals. We therefore sought to compare the algorithm-induced errors of the jet structures with the inherent temporal noise of the HRI$_{\text{EUV}}$ instrument itself. The temporal noise on repeated HRI$_{\text{EUV}}$ pixel values is typically dominated by the sensor read noise and the photon shot noise according to the formula \citep{euidatarelease6}:
\begin{equation}
s^2 = r^2 + I * t * a/n
,\end{equation}
where $s$ is the uncertainty on measured value, $r$ is the readout noise (2 DN), $I$ is the measured intensity (DN/s), $t$ is the exposure time (s), $a$ is the photon to DN conversion factor (6.85 DN/photon), and $n$ is the sample size (number of pixels over which the intensity is averaged). As shown in Fig.~\ref{fig11}, we compared the algorithm-induced errors  at compression ratios of 137× and 495× with the temporal noise of the HRI$_{\text{EUV}}$ instrument. For the intensity distribution across the jet shown, the peak intensities in the reconstructed and original images were 5249:5211 and 5152:5211, indicating errors of 38 DN/s (137×) and -59 DN/s (495×), respectively. For $I$ = 5211 DN/s with an exposure time of $t$ = 2.0 s, the uncertainty on the measured value is 267 DN. This demonstrates that even at a compression ratio of 495x, the error in the peak intensity of the jet in the reconstructed image due to the algorithm remains significantly smaller than the measurement uncertainty caused by the HRI$_{\text{EUV}}$ temporal noise.

Our comprehensive analysis indicates that for this set of ultra-high-resolution HRI$_{\text{EUV}}$ observations near perihelion, SolarZip can effectively achieve compression ratios of several hundred times without affecting scientific analysis. Specifically, for high-contrast dynamic structures represented by solar jets, compression ratios of around 500× can be achieved (corresponding to a 2e-2 error bound). For quiescent prominences, even higher compression ratios of approximately 800× are possible (corresping to a 3e-2 error bound).

\section{Conclusions}
\label{sec:6}
This paper presents SolarZip, an efficient and adaptive compression and evaluation framework specifically designed for solar EUV data from solar missions. To our knowledge, this is the first study to systematically apply and analyze advanced error-bounded lossy compressors (SZ, ZFP, and SPERR) on solar EUV data, demonstrating their significant advantages over traditional methods. However, their inherent limitations motivated us to design an adaptive hybrid compression strategy that dynamically selects optimal decorrelation models based on observational scenarios, coupled with optimized interpolation predictors to enhance compression quality. SolarZip achieved unprecedented compression ratios of up to 800× for EUI/FSI data and 500× for  EUI/HRI$_{\text{EUV}}$ data, surpassing traditional algorithms by 3-50×. Our comprehensive two-stage evaluation framework ensures that compressed data remains suitable for critical scientific research by integrating strict error control with downstream scientific workflows. The simulation experiments based on Solar Orbiter hardware conditions confirm that SolarZip can reduce data transmission time by a factor of 270, addressing a critical bottleneck in deep space solar missions.

The test data used in this work are publicly available EUI level-1 data that have already undergone on-board "lossy-high quality" compression processing. As demonstrated in Appendix \ref{appendixA}, these prior processing steps inherently limit the compression performance of SolarZip, as the data have already been subject to lossy compression, dynamic range reduction, and RICE encoding. Our comparative analysis across different on-board compression modes (Table~\ref{tab2}) reveals that SolarZip is expected to achieve 15–50\% higher compression ratios when applied to less processed data, such as those compressed on-board using lossless modes. The mathematical analysis in Appendix \ref{appendixC} provides a theoretical justification that our performance evaluation on level-1 data remains scientifically valid and representative, as the preprocessing operations preserve the essential solar physical patterns that determine compression effectiveness. This suggests that  SolarZip's performance on raw or minimally processed solar data would be even more effective, going beyond the already substantial improvements demonstrated in this work.

Future works will focus on algorithmic enhancements tailored to the specific properties of solar data. For example, we aim to introduce a region of interest (ROI) approach, enabling the application of tighter error constraints in scientifically critical regions to achieve superior compression performance. Moreover, the versatility of our compression framework suggests its potential applicability to other types of astronomical datasets, such as radio imaging observations, which represents a promising avenue for future exploration.

\section{Data availability}
Solar Orbiter data is publicly available through the Solar Orbiter Archive\footnote{https://soar.esac.esa.int/soar/}.The sample data used to generate the figures in this work are publicly available at the following link: SolarZip-TestData\footnote{https://github.com/CapitalLiu/SolarZip-TestData}. The dataset includes the raw data and the decompressed reconstructed data for each figure. The source code will be made publicly available at SolarZip\footnote{https://github.com/hpdps-group/SolarZip}.

\begin{acknowledgements}
We thank the anonymous reviewer for the constructive comments, which undoubtedly significantly improved the scientific quality and readability of the manuscript. We also thank the Solar Orbiter/EUI team at the Royal Observatory of Belgium for their generous assistance, particularly David Berghmans and Emil Kraaikamp for their help with the use and understanding of EUI data.
Solar Orbiter is a space mission of international collaboration between ESA and NASA, operated by ESA. The EUI instrument was built by CSL, IAS, MPS, MSSL/UCL, PMOD/WRC, ROB, LCF/IO with funding from the Belgian Federal Science Policy Office (BELPSO); the Centre National d'Etudes Spatiales (CNES); the UK Space Agency (UKSA); the Bundesministerium f\"ur Wirtschaft und Energie (BMWi) through the Deutsches Zentrum f\"ur Luft- und Raumfahrt (DLR); and the Swiss Space Office (SSO).
D.T. and G.T would like to acknowledge support from the National Natural Science Foundation of China (Grant Nos. 62032023, and T2125013) and the Innovation Funding of ICT, CAS (Grant No. E461050). B.Z. was supported by the NSFC Fund (042274216). The AIP team was supported by the German Space Agency (DLR), grant number \mbox{50 OT 2304}. This research used the SunPy \citep{sunpy_community2020,Mumford2020} and NicePlots\footnote{https://github.com/mdolab/niceplots} software package to present the observation results. 
\end{acknowledgements}

\bibliographystyle{aa} % style aa.bst
\bibliography{ref} % your references Yourfile.bib

\begin{appendix}
\section{Data and Supplementary Experiments}
\label{appendixA}

The EUI data undergo a comprehensive multi-stage processing pipeline, beginning with significant on-board data reduction before the generation of scientific data products on the ground. The process is initiated on-board with the simultaneous acquisition of two 12-bit images: a high gain (HG) channel optimized for faint features and a low gain (LG) channel for bright structures. These images first undergo an initial calibration to correct for instrumental artifacts, most notably a strong, systematic four-column banding noise. Following this, the calibrated HG and LG images are merged to create a single 15-bit "combined gain" image with an extended dynamic range. This 15-bit image is then subjected to an integer square root recoding, converting it to an 8-bit format. This recoding serves the dual purpose of suppressing shot noise and preparing the data for the on-board compression hardware, which requires an eight-bit input. The final on-board step is a lossy wavelet-based compression. Once transmitted to the ground, these compressed data packets are decompressed, and the recoding is reversed to reconstruct the image's bit depth, producing level-1 (L1) data. These L1 files are packaged using a lossless RICE tile compression. Subsequent processing to create level-2 (L2) data involves further scientific calibrations, such as optical distortion correction for FSI, which necessitates data resampling. The L2 files are also tile-compressed, although this can introduce minor quantization artifacts due to the scaling of floating-point data prior to compression.

\begin{figure}[htbp]
\centering
    \includegraphics[width=\linewidth]{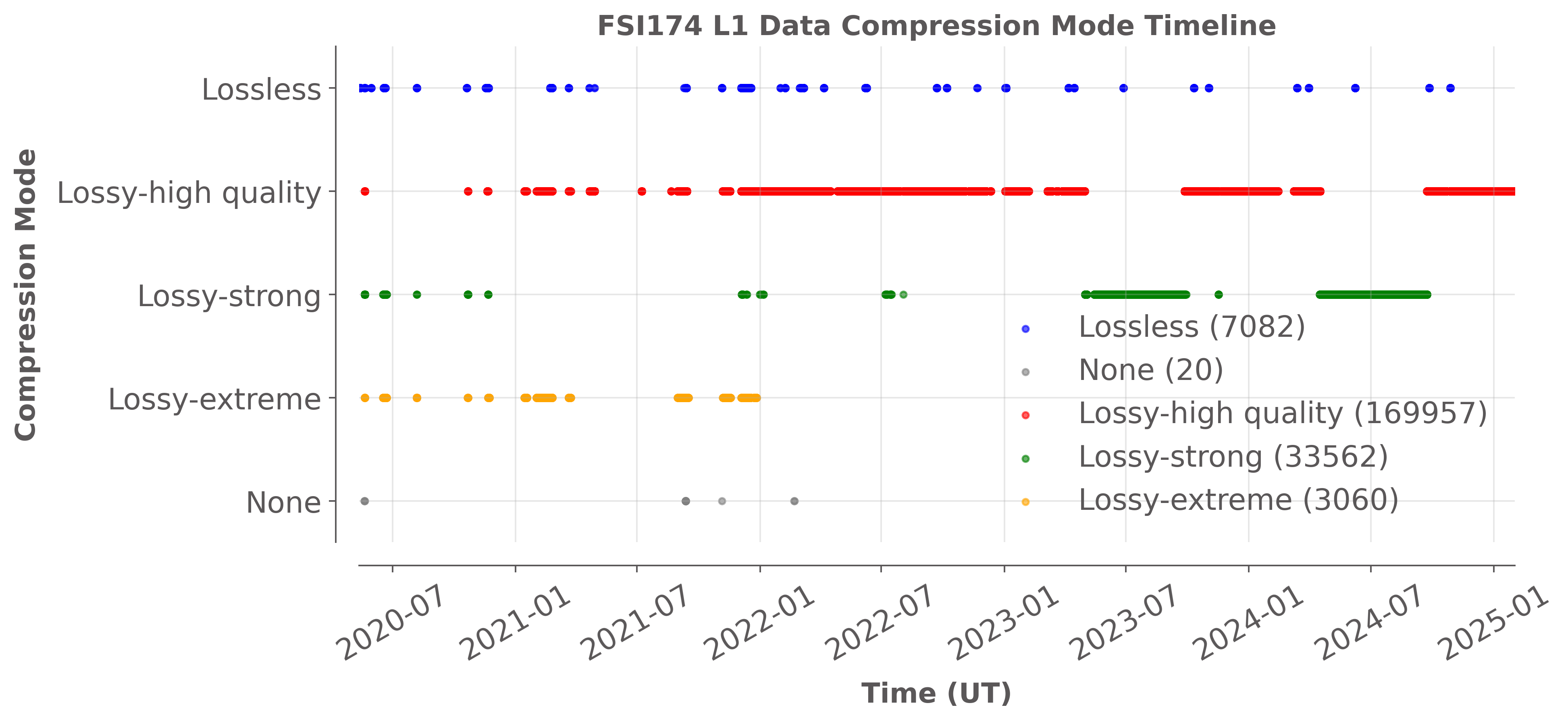}
\caption{Timeline distribution of FSI 174 A L1 data compression modes in EUI Data Release 6.0.}
\label{figB1}
\end{figure}

On-board compression on EUI includes four quality modes: lossless, lossy-high quality, lossy-strong, and lossy-extreme. Ideally, the best case for evaluating our SolarZip algorithm would be to use EUI data compressed in the lossless mode. However, we ultimately chose data compressed using the "Lossy--high quality" on-board mode as the primary dataset for our experiments, since this mode accounts for the largest proportion of the data and spans the widest time range (see the accompanying analysis of FSI-L1 compression mode distribution and temporal coverage for details).

In addition, we sought to address a key question concerning whether the on-board compression mode affects the performance of downstream compression algorithms such as SolarZip. To investigate this possibility, we conducted additional experiments. As shown in Table~\ref{tab2}, We traversed all possible data (data with four different compression modes in one hour, totaling 15 groups of 60 files), under different on-board compression modes to evaluate the resulting compression ratios. The results clearly indicate that the on-board compression mode does have a significant impact on SolarZip’s performance. The more aggressively the data was compressed on-board by WICOM, the lower the additional compression ratio achieved by SolarZip.

\begin{table}[htbp]
\centering
\caption{Average compression ratios (15 groups) for L1 data: which were compressed by four on-board modes and then recompressed by SolarZip.}
\label{tab2}
\footnotesize
\begin{tabular*}{1\columnwidth}{@{\extracolsep{\fill}}lcccc@{}}
\toprule
\textbf{Mode} & \textbf{Lossless} & \textbf{Lossy-HQ} & \textbf{Lossy-strong} & \textbf{Lossy-extreme} \\
\midrule
$1e-4$ & 56.33 & 45.61 & 32.92 & 25.34 \\
$1e-3$ & 198.15 & 152.66 & 131.84 & 124.17 \\
$1e-2$ & 659.24 & 574.83 & 506.72 & 395.25 \\
\bottomrule
\end{tabular*}
\end{table}

It is worth noting that the EUI team retains a very limited set of data that is both lossless and unrecoded. However, these images use single-gain channels, which differs from the combined-gain images that make up the vast majority of observations. As this dataset is not publicly available, it was not included in our tests. Nonetheless, our comparison across different L1 compression modes already provides sufficient evidence for our conclusion: when applied to less compressed (more original) data, SolarZip consistently achieves higher compression ratios. This further demonstrates the broad applicability and effectiveness of SolarZip.

\section{Lossy Compression Technique}
\label{appendixb}
\subsection{Advanced lossy compressors}
\label{appendixa}

Appendix B provides a detailed description of the four advanced error-bounded lossy compression algorithms employed for evaluation and comparison in this study. We first present their workflows and then highlight their respective characteristics.

SZ2 \citep{liang2018efficient} refers to the second-generation SZ compression algorithm, Which is a prediction-based error-bounded lossy compressor.\\
Workflow: SZ2's compression has four main steps. Firstly, it divides raw data to be compressed into small blocks. For each of these blocks, a separate prediction function is generated. Secondly, SZ2 quantizes the data with the specified error bound. Thirdly, it employs Huffman encoding to encode the quantization index. Finally, lossless compression methods are used to further improve the compression ratio.\\
Insight: The SZ2 compression process is simple and efficient, achieving notable compression performance and speed on most scientific datasets. However, due to its block-wise prediction approach, its effectiveness is limited for high-dimensional and nonlinear data.

\label{sec2.2}
SZ3 \citep{liang2022sz3,zhao2021optimizing} uses a modular approach to compress the data. In fact, it is not only a compression software but also a flexible framework allowing users to customize specific copmression pipelines according to their datasets or use cases.\\
Workflow: The SZ3 compression pipeline is composed of five stages: preprocessing, prediction, quantization, variablelength encoding, and lossless compression. The preprocessing stage starts by transforming and shaping the raw data to make it easier to compress. The second stage is prediction. For different domain datasets, SZ developers have developed many predictors, including Lorenzo, linear regression, and dynamic spline interpolation. Thirdly, the error produced by the predictor is quantized. Fourthly, the quantized error data is encoded, shrinking its size. Fifthly, the now encoded data is losslessly compressed, reducing the size even further.\\
Insight: With an advanced predictive model and a flexible modular design, SZ3 significantly enhances compression performance and adaptability. However, these improvements come with higher computational costs.

ZFP \citep{lindstrom2014fixed,diffenderfer2019error} is a transform-based  error-bounded lossy compressor.\\
Workflow: ZFP splits the whole dataset into many fixed-size blocks (e.g., 4×4×4 for a 3D dataset) These blocks are then individually compressed. The ZFP compressor executes four steps in each block. The first step align the value in block to a common exponent and convert the floating-point values to a fixed-point  repressention. The next step uses orthogonal block transform to decorrelate data. Thirdly, it orders the transform coefficient by expected magnitude. Finally, it encodes the coefficients to reduce data size.\\
Insight: ZFP generally features high compression and decompression performance because of the performance optimization strategies in its implementation. However, as a block-wise compressor, ZFP faces the same limitations as SZ2.

SPERR  is a transform-based lossy compressor based on the CDF9/7 discrete wavelet transform and SPECK encoding algorithm \citep{li2023lossy}. \\
Workflow: Compression pipeline of SPERR includes four stages: (1)CDF9/7 wavelet transform; (2) SPECK lossy encoding of wavelet coefficients; (3) outlier encoding (only in error-bounding mode); (4) zstd postprocessing of compressed data (optional).\\
Insight: SPERR's advantage is that the hierarchical multidimension DWT in SPERR can effectively capture the relevance between data points, which brings a high compression ratio after the SPECK encoding. One limitation of SPERR is that the wavelet transform and the SPECK encoding processes have high computational costs; hence, its (sequential) execution speed is  low, typically around 30\% of SZ3.

\subsection{Decorrelation module}

\begin{figure}[htbp]
\centering
    \includegraphics[width=\linewidth]{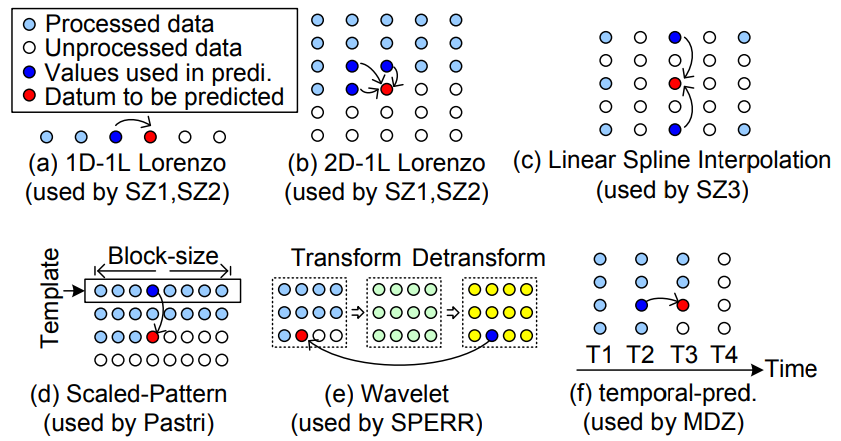}
\caption{Categories of six types of decorrelation modules. The Lorenzo predictor is used in SZ2, spline interpolation is adopted in SZ3, and wavelet transform is applied in SPERR.}
\label{figB1}
\end{figure}

The decorrelation module is a critical component in lossy compressors, as it significantly affects the compressor’s performance. In general, the more accurately a predictor can model the inherent patterns in the data, the better the decorrelation it achieves. Data that has been decorrelated is more amenable to compression. Below, we introduce several mainstream predictors in detail.

The earliest proposed predictor is the Lorenzo predictor, which uses adjacent data points to predict the next one. For example, in the 1D case shown in Figure~\ref{figB1}, the value of the next point is always predicted using the value of the point to its left. This can be viewed as a form of extrapolative prediction. Later, spline interpolation was introduced as an alternative, which can be considered an interpolative approach. This method first predicts long-range points and then uses multiple points to estimate the intermediate values. Its advantage lies in its ability to capture more high-dimensional and global information.

\begin{figure}
\centering
\includegraphics[width=\linewidth]{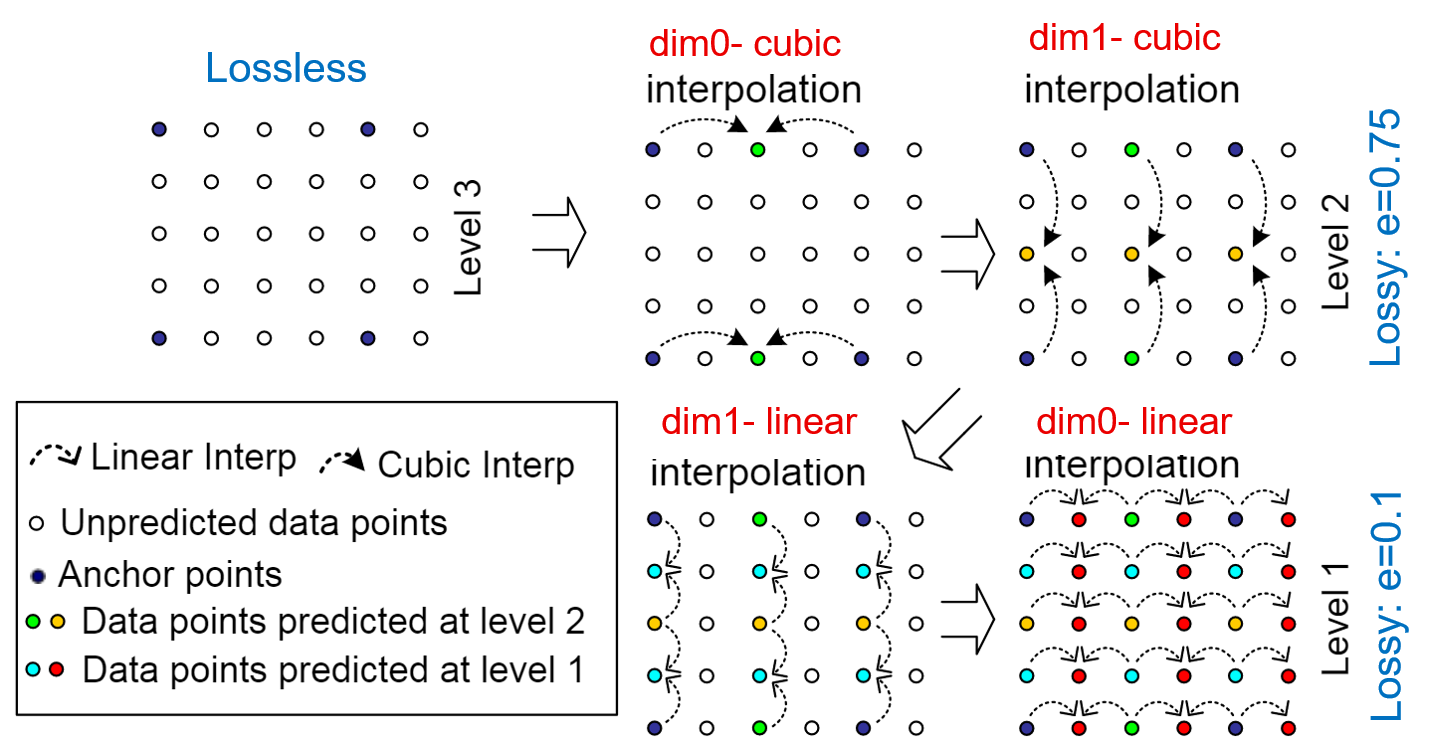}
\caption{Level-wise anchor points based dynamic spline interpolation. In 2D data, there are two interpolation directions, and each interpolation step can utilize either a linear or cubic interpolation function. The algorithm automatically optimizes the selection process. It is evident that at the more critical level 2, a smaller error bound is assigned, which further enhances the accuracy of the interpolation prediction.}
\label{fig:interpolation}
\end{figure}

Unlike numerical prediction methods, transform-based approaches operate in the data domain. An example is the wavelet transform used in SPERR. Wavelet transforms convert the original data into a domain that is more favorable for compression. In this domain, the data is represented as coefficients with varying levels of significance. More information is concentrated in the more important coefficients, allowing the subsequent compression algorithm to selectively retain these while discarding or coarsely compressing the less significant ones. Transform-based approaches often yield better compression quality, but they come at a higher computational cost.

\subsection{Anchor points interpolations with error bound auto-tuning}
\label{appendixb3}
An issue is that classical spline interpolation applies the same error bound to all prediction levels, which does not fully account for their relative importance. As shown in the Fig. \ref{fig:interpolation}, the predicted data points in level 1 participate in five subsequent prediction steps. Consequently, earlier-predicted data points should be considered more important. 

First, we apply different interpolation methods at different levels of interpolation prediction. Specifically, the interpolation types include linear interpolation and cubic interpolation. In our 2D data, the interpolation process is essentially carried out through multiple 1D interpolation operations. Since there are two dimensions (dim0 and dim1), two distinct interpolation orderings are possible, and we select the optimal interpolation sequence.

Next, we set different error bounds at different levels of interpolation (as opposed to the unified error bounds used in SZ3). In our 2D data, 75\% of the data points fall within the lowest interpolation level (level 1), which are predicted by higher-level reconstructed data points, while the remaining 25\% of the data points are predicted at higher levels. Therefore, setting smaller error bounds at higher levels helps ensure overall prediction accuracy, thus improving compression quality. The level-wise error bounds, \(e_l\), are dynamically adjusted based on equation \ref{level error}. The parameters \(\alpha\) and \(\beta\) are introduced, where \(e\) represents the global error bound set by the user. We perform offline testing with parameter sets \(\alpha = \{1, 1.5, 2\}\) and \(\beta = \{2, 3, 4\}\), comparing the bit-rate and PSNR values across different parameter configurations. Ultimately, we selected \(\alpha = 1.5\) and \(\beta = 4\) as our optimal parameters.
\begin{equation}
    \text{ }e_l = \frac{e}{ min(\alpha^{l-1}, \beta)} \quad (\alpha \geq 1 \text{ and } \beta \geq 1)
    \label{level error}
.\end{equation}

\section{Mathematical proof: Reliability of compression performance evaluation on EUI data products}
\label{appendixC}

\subsection{Theorem Statement}

Theorem: \textit{Let $\mathcal{D}_0, \mathcal{D}_2$ represent raw and level-2 solar EUI data respectively, where $\mathcal{D}_2$ results from sequential preprocessing operations including calibration $\mathcal{C}$, on-board compression $\mathcal{P}$, and RICE encoding $\mathcal{R}$. For SolarZip ($\mathcal{A}_{SZ}$) employing spline interpolation prediction, the compression performance metrics (compression ratio, $CR,$  and distortion measures RMSE and PSNR) evaluated on $\mathcal{D}_2$ remain statistically equivalent to those evaluated on $\mathcal{D}_0$, as preprocessing preserves the underlying solar physical patterns while reducing noise and instrumental artifacts.}

\subsection{Mathematical Framework}

For solar imagery, we decompose the data into physical components,
\begin{equation}
\mathcal{D}_0 = \mathcal{S}_0 + \mathcal{N}_0 + \mathcal{A}_0
,\end{equation}
where $\mathcal{S}_0$ represents underlying solar physical patterns, $\mathcal{N}_0$ is random noise, and $\mathcal{A}_0$ denotes instrumental artifacts.

The preprocessing sequence preserves physical patterns while reducing noise,
\begin{equation}
\mathcal{D}_2 = \mathcal{R}(\mathcal{P}(\mathcal{C}(\mathcal{D}_0))) = \mathcal{S}_0 + \epsilon\mathcal{S}_0 + \mathcal{N}_2 + \mathcal{A}_2
,\end{equation}
where $|\epsilon| \ll 1$, $\|\mathcal{N}_2\|_2 \leq \|\mathcal{N}_0\|_2$, and $\|\mathcal{A}_2\|_2 \leq \|\mathcal{A}_0\|_2$.

\subsection{Spline interpolation analysis for SolarZip}

SolarZip employs B-spline interpolation for predictive compression. The B-spline basis functions of degree $p$:
\begin{equation}
B_{i,0}(t) = \begin{cases} 1 & \text{if } t_i \leq t < t_{i+1}, \\ 0 & \text{otherwise}, \end{cases}
\end{equation}

\begin{equation}
B_{i,p}(t) = \frac{t - t_i}{t_{i+p} - t_i} B_{i,p-1}(t) + \frac{t_{i+p+1} - t}{t_{i+p+1} - t_{i+1}} B_{i+1,p-1}(t)
.\end{equation}

The spline approximation for solar data,
\begin{equation}
\mathcal{S}(x,y) \approx \sum_{i,j} c_{ij} B_i^p(x) B_j^q(y)
.\end{equation}

Lemma: \textit{Solar physical patterns exhibit smooth spatial variations amenable to spline representation.}

Proof: Solar magnetic field configurations follow MHD equations, yielding smooth solutions. The approximation error for B-splines of degree $p$:
\begin{equation}
\|\mathcal{S} - \sum_{i,j} c_{ij} B_i^p(x) B_j^q(y)\|_\infty \leq C h^{p+1} \|\mathcal{S}^{(p+1)}\|_\infty
.\end{equation}

Since preprocessing maintains smoothness: $\|\mathcal{S}_2^{(p+1)} - \mathcal{S}_0^{(p+1)}\|_\infty \leq \epsilon_s$, the spline approximation quality remains consistent. $\square$

\subsection{Compression Performance Invariance}

The prediction error using spline interpolation:
\begin{equation}
e[n] = x[n] - \sum_{i=0}^{N} c_i B_i^p(n)
.\end{equation}

For optimal coefficients minimizing $\|e\|_2^2$, we solve:
\begin{equation}
\mathbf{G}\mathbf{c} = \mathbf{b}
,\end{equation}
where $G_{ij} = \langle B_i^p, B_j^p \rangle$ and $b_j = \langle x, B_j^p \rangle$.

Since preprocessing preserves smooth solar patterns: $\|\mathbf{b}_2 - \mathbf{b}_0\|_2 \leq \epsilon_b$, the prediction quality remains consistent:
\begin{equation}
\|\mathbf{e}_2\|_2^2 - \|\mathbf{e}_0\|_2^2 \leq 2\epsilon_b \|\mathbf{c}\|_2 + \epsilon_b^2
\end{equation}

\subsection{Performance bounds}

For the compression ratio:
\begin{equation}
\left|\frac{CR_2 - CR_0}{CR_0}\right| \leq \frac{\epsilon_s}{\text{S/N}_{pattern}}
.\end{equation}

For distortion measures (RMSE):
\begin{equation}
\frac{|\text{RMSE}_2 - \text{RMSE}_0|}{\text{RMSE}_0} \leq \frac{\epsilon_s}{\text{RMSE}_0}
,\end{equation}

where $\text{S/N}_{pattern} = \frac{\|\mathcal{S}_0\|_2^2}{\|\mathcal{N}_0 + \mathcal{A}_0\|_2^2}$ is the pattern-to-noise ratio.

\subsection{Conclusion}

The mathematical analysis demonstrates that compression performance evaluation on L1/L2 solar imaging data maintains full scientific validity. The preprocessing operations preserve essential physical patterns that determine SolarZip's effectiveness while reducing detrimental noise and artifacts.

Key Results:
\begin{itemize}
\item Spline interpolation accuracy is maintained due to preserved spatial smoothness;
\item Compression ratios and distortion measures exhibit bounded, negligible deviations;
\item Statistical significance of evaluations is preserved or enhanced.
\end{itemize}

Therefore, L1/L2 imaging data (under difference compression mode) provides a reliable basis for evaluating SolarZip compression performance on solar imagery data.

\end{appendix}
\end{document}